\documentclass[aps,twocolumn,superscriptaddress,nofootinbib,prb]{revtex4-2}

\usepackage{graphicx}
\usepackage{placeins} 
\usepackage{amssymb,amsmath,amsfonts,bm}
\usepackage{subcaption}
\usepackage[normalem]{ulem}
\usepackage{hyperref}
\usepackage[dvipsnames]{xcolor}
\hypersetup{urlcolor=RoyalBlue, colorlinks=true,citecolor=RoyalBlue,linkcolor=RoyalBlue} 
\usepackage{physics}
\usepackage[final]{showkeys}
\usepackage{float}
\usepackage{comment}
\usepackage{amsthm}
\usepackage{bbold}

\newtheorem{theorem}{Theorem}

\begin{document}
\title {Quantum many-body scars and zero modes in the spin-1 Kitaev chain}

\author{Ayush Sharma}
\email{ayush.sharma@tifr.res.in}
\affiliation{Department of Theoretical Physics, Tata Institute of Fundamental Research, Homi Bhabha Road, Navy Nagar, Mumbai 400005, India}

\author{Jay Pandey}
\email{jay.pandey@tifr.res.in}

\affiliation{Department of Theoretical Physics, Tata Institute of Fundamental Research, Homi Bhabha Road, Navy Nagar, Mumbai 400005, India}


\begin{abstract}

We study the spin-1 Kitaev chain and, by analyzing its Hilbert-space connectivity in the Cartesian basis, show that it hosts an exponentially large number of exact zero-energy modes protected by chiral symmetry. For even-length periodic chains, their count is lower-bounded by the Jacobsthal number $(2^{L+1}+1)/3$, where $L$ is the chain length. We then construct an exponentially large number ($2^{L/2}$) of zero modes with an analytic expression that have a maximally mixed density matrix for a contiguous bipartition using the operator-state mapping. We further identify an embedded tight-binding structure governing defect dynamics that supports PXP-like quantum many-body scars (QMBS) in multiple sectors previously not reported. The associated forward-scattering approximation (FSA) Hamiltonian takes a universal form across the scarred sectors studied. A simple physical picture of defects propagating between walls provides an intuitive description of these dynamics and a structural guide to identifying sectors with fidelity revivals, while its limitations point to additional scarring mechanisms. 

\end{abstract}

\maketitle

\section{Introduction}

When a quantum many-body system is prepared in a state far from equilibrium, its subsequent dynamics generally drive it towards thermal equilibrium. This behavior is captured by the eigenstate thermalization hypothesis (ETH)~\cite{Srednicki:1994mfb, Deutsch:1991msp}, which explains how the interactions between different degrees of freedom effectively allow the system to act as its own reservoir, leading to the loss of memory of the initial state. Two well-established exceptions to the ETH paradigm are integrable systems and many-body localized (MBL) systems~\cite{MBL_review, MBL_Apal, BASKO20061126}.

More recently, a distinct mechanism for weakly breaking ergodicity has been identified~\cite{Bernien2017}. This phenomenon is associated with the presence of a set of atypical, non-thermal eigenstates embedded in an otherwise thermal spectrum, dubbed quantum many-body scars (QMBS)~\cite{Turner_2018_PXP}. Since their discovery, QMBS have been identified in a wide variety of models from spin systems~\cite{eta_pairing, AKLT_model_1, AKLT_model_2, DWC_model, spin1xy, Unified_mark_lin_motrunich, eta_pairing_extendedhubbardmodel}, quantum hall system~\cite{Sanjay_scar_landau_levels}, lattice gauge theories\cite{scars_in_abelian_LGT, quantum_link_dimer_model_Arnab, scars_schwinger_model, Non_mesonic_scars_1D_LGT}, frustrated systems ~\cite{scars_frustration_3coloured, Arnab_scars_frustration}, and various versions of the PXP model~\cite{Exact_higher_spin_PXP, stability_2D_pxp, scars_2D_pxp, scars_beyond_pxp, Driven_PXP_model}(for a detailed review of the subject, we refer the reader to~\cite{Moudgalya_2022, Serbyn2021_review}). Conserved quantities partition the Hilbert space into disconnected sectors, within which thermalization occurs in certain sectors while it may not happen in others. This raises the question of how the sector structure shapes QMBS and their observable signatures.

The spin-$1$ Kitaev chain provides a useful setting for addressing this question. Its alternating bond-dependent interactions admit an extensive set of mutually commuting $\mathbb{Z}_2$ bond charges, which divide the Hilbert space into exponentially many sectors~\cite{Spin-1_Kitaev_model_Dhar}. Previous works established an exact mapping of one sector of the chain onto the PXP model and identified the associated scarring dynamics~\cite{Spin-1_Kitaev_model_Dhar,You_scars_in_Kitaev_model}. Subsequent studies have reported scarring in additional sectors~\cite{You_scars_in_Kitaev_model,Balram_Kitaev_scars}.

These results motivate a systematic investigation of how the bond-charge configuration shapes the spectrum and dynamics beyond the previously studied examples. We therefore study the $S=1$ Kitaev chain. Exact zero modes provide an analytically accessible starting point for understanding atypical eigenstates within conserved sectors. For a subset of them, we examine their wavefunction across different sectors and their explicit construction, motivated by antipodally paired eigenstates with unusual correlations~\cite{Volume_law_scar_PXP,Exact_volume_law_zero_modes}. We then turn to sectors labeled by the conserved $\mathbb{Z}_2$ bond charges that exhibit scarred dynamics, examining how the bond-charge configuration controls whether scarring emerges, and identifying a common dynamical description shared by several classes of such sectors.

Focusing first on its exact zero-energy structure, the clean chain hosts exponentially many zero modes, including zero modes associated with space-group symmetries~\cite{zm_count_buijsman}
within individual sectors. These symmetries depend on the system size $L$ and individual sectors, complicating a complete characterization of the zero-energy manifold. To isolate the zero modes not protected by the space group symmetries, we introduce random bond disorder that breaks these spatial symmetries while preserving chiral symmetry. We establish that exponentially many exact zero modes survive in total, with their number bounded from below in each sector, with a sector-dependent bound called the chiral index of that sector. We find an analytical expression for the sum of these chiral indices, called the total chiral index, which is found to be equal to the Jacobsthal numbers~\cite{weisstein_jacobsthal} $J_{L+1} = (2^{L+1}+1)/3$. In the presence of weak bond disorder, the numerically obtained number of zero modes saturates this bound exactly for all system sizes studied (see Sec.~\ref{sec:ZeroModeCounting}). We then derive exact analytic expressions for the wavefunctions of antipodally correlated zero modes in each sector invariant under translation by $L/2$, using an approach that provides a general framework for constructing such zero modes beyond the specific families considered in Refs.~\cite{Volume_law_scar_PXP,Exact_volume_law_zero_modes}. Although applied here to a chiral-symmetric model, our approach can be extended beyond chiral-symmetric models; see the discussion for details. Across all the symmetry sectors, we find $2^{L/2}$ such states exist for an even-length $L$ chain with periodic boundary conditions (PBC). 

We next investigate the scarred dynamics within different sectors of the chain. Previous studies \cite{You_scars_in_Kitaev_model, Zhang_scars_Kitaev_single_ion_anisotropy, Balram_Kitaev_scars} have identified certain sectors of this model exhibiting scarred dynamics with phenomenology similar to that of the PXP model~\cite{Turner_2018_PXP}. Rationale for these, in observed sectors, is provided by our observation of embedded tight-binding structure in the sectors; using this, we extend these results by showing that QMBS occur in a broader class of sectors than previously reported, particularly those characterized by periodic bond-charge configurations and corresponding ``defect-free'' vacuum states (see Sec.~\eqref{sec:SectorResolvedApproximateScarStudy} for details). We analyze these sectors systematically and show that the associated scar states are described by the forward scattering approximation (FSA)~\cite{Turner_2018_PXP, Turner2018}. The FSA Hamiltonian takes the same form across all the scarred sectors, suggesting a universal mechanism underlying the emergence of these states. 

We illustrate this through an intuitive physical picture, which organizes allowed processes and offers a heuristic of which sectors possess scarring dynamics. While the FSA provides a common description of the scarring phenomenon, the resulting revival dynamics are not uniform across sectors: some exhibit strong fidelity revivals, whereas others show substantially weaker ones. We further argue that certain sectors cannot harbor QMBS due to this mechanism from the defined ``defect-free" states and provide numerical support for the same. This does not rule out the scarring dynamics due to some other mechanisms in these sectors.

The rest of this paper is organized as follows. Sec.~\ref{sec:Model} introduces the model and its symmetries, and Sec.~\ref{sec:ZeroModeCounting} presents the counting of zero modes using the transfer-matrix formalism. In Sec.~\ref{sec:VolumeLawEntangledScars}, we construct antipodally correlated zero modes using operator-state mapping~\cite{Preskill1998-aq}, offering an alternative to the construction in Ref.~\cite{Exact_volume_law_zero_modes}. Section~\ref{sec:SectorResolvedApproximateScarStudy} studies approximate scars across different classes of symmetry sectors, and Sec.~\ref{sec:NonScarredSectors} analyzes the sectors in which the chosen protocol fails to produce scars. Finally, Sec.~\ref{sec:Discussion} contains a discussion on all of our findings.

\section{Model and its symmetries}
\label{sec:Model}

\begin{figure}
    \centering
    \includegraphics[width=\linewidth]{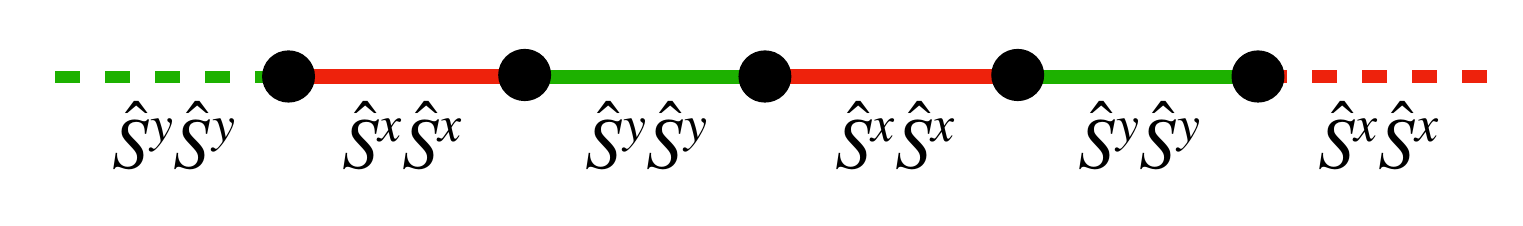}
    \caption{A schematic of the Kitaev chain. Black sites host the $S=1$ spins. $S^{x}S^x$ and $S^{y}S^y$ bond couplings are shown in red and green respectively.}
    \label{fig:KitaevSchematic}
\end{figure}

The spin-1 Kitaev chain is described by
\begin{equation}\label{eq:KitaevHamiltonian}
H=\sum_{i}\Bigl(J_xS^{x}_{2i-1}S^{x}_{2i}+J_yS^{y}_{2i}S^{y}_{2i+1}\Bigr),
\end{equation}
where alternating nearest-neighbor bonds host Ising-like interactions of $x$ and $y$ type. Here, $S^{\alpha}$, with $\alpha={x,y,z}$, denote the spin-1 operators. Throughout this work, we impose PBC and use even system sizes, unless specified. A schematic of $H$ is shown in Fig.~\ref{fig:KitaevSchematic}. This model arises as a one-dimensional limit of the Kitaev honeycomb model~\cite{KITAEV_2006_2} by eliminating one bond species, say the $S^{z}S^{z}$ interaction. This fragments the two-dimensional system into a set of decoupled chains governed by Eq.~\eqref{eq:KitaevHamiltonian}. Unless otherwise stated, we consider homogeneous couplings, $J_x=J_y$. The chiral index derived in Sec.~\ref{sec:ZeroModeCounting}, however, also holds for $J_x\neq J_y$. 

To identify the symmetries of Eq.~\eqref{eq:KitaevHamiltonian}, we use the local parity operators.
\begin{equation}\label{eq:ParityOp}
\mathcal{P}^{\alpha}_{i}=e^{i\pi S^{\alpha}_{i}}.
\end{equation}
As shown in Appendix~\ref{app:Directorbasis}, they satisfy
\begin{equation}\label{eq:parity_relations}
[\mathcal{P}^{\alpha}_{i},S^{\alpha}_{j}]=0 \quad (i \neq j),
\qquad
\{\mathcal{P}^{\alpha}_{i},S^{\beta}_{i}\}=0
\quad (\alpha\neq\beta).
\end{equation}
The alternating $S^xS^x$ and $S^yS^y$ couplings then imply an extensive set of conserved bond-parity operators,
\begin{align}
\mathcal{B}^{y}_{2i-1} &= \mathcal{P}^{y}_{2i-1}\mathcal{P}^{y}_{2i}, \\
\mathcal{B}^{x}_{2i}   &= \mathcal{P}^{x}_{2i}\mathcal{P}^{x}_{2i+1}.
\label{eq:Bondoperators}
\end{align}
which commute with the Hamiltonian and label its fragmented sectors.
These operators are conserved because a simultaneous $\pi$ rotation about the $y$ axis on the bond $(2i-1, 2i)$ leaves the corresponding $S^{y}S^{y}$ interaction invariant. An analogous argument applies to the $x$ bonds.

It is convenient to perform the unitary transformation on the even sites introduced in Ref.~\cite{Spin-1_Kitaev_model_Dhar},
\begin{equation*}
S^{x}_{2i} \to S^{y}_{2i}, \quad S^{y}_{2i} \to S^{x}_{2i}, \quad S^{z}_{2i} \to -S^{z}_{2i},
\end{equation*}
under which the Hamiltonian in Eq.~\eqref{eq:KitaevHamiltonian} is mapped to the translationally invariant form
\begin{equation}\label{eq:TranslationalinvariantKitaev}
H = \sum_{i} S^{x}_{i}S^{y}_{i+1},
\end{equation}
while the conserved bond operators take the translationally invariant form
\begin{equation}\label{eq:Translationinvariantbondoperators}
\mathcal{B}_{j} = \mathcal{P}^{y}_{j}\mathcal{P}^{x}_{j+1}.
\end{equation}

Since $\mathcal{B}_j^2 = \mathbb{1}$, the eigenvalues of each $\mathcal{B}_j$ are restricted to $\pm 1$. For a chain of $L$ sites, the $3^L$- dimensional Hilbert space therefore decomposes into $2^L$ symmetry sectors, one for each joint eigenspace of the mutually commuting and independent bond operators. Each sector is labeled by a bond configuration $\{w_1, w_2, \dots, w_L\}$, where $w_j = \pm 1$ denotes the eigenvalue of $\mathcal{B}_j$.

Before proceeding further, we introduce a convenient local basis, which we will use throughout this work, referred to as the Cartesian basis,
\begin{align}\label{eq:Directorbasis}
|x\rangle &= \dfrac{|-1\rangle - |1\rangle}{\sqrt{2}},  \
|y\rangle &= \dfrac{i\left(|-1\rangle + |1\rangle\right)}{\sqrt{2}}, \
|z\rangle &= |0\rangle,
\end{align}
where $|-1\rangle$, $|0\rangle$, and $|1\rangle$ denote the eigenstates of the $S^{z}$ operator with eigenvalues $-1$, $0$, and $1$, respectively.

\section{Zero modes from imbalance in Hilbert space}\label{sec:ZeroModeCounting}

We now study the exact zero modes of the Hamiltonian due to chiral symmetry in Eq.~\eqref{eq:TranslationalinvariantKitaev}, excluding those protected by the space group symmetries, whose count we refer to as the chiral index across different sectors. For chains of even length $L$, the model possesses a chiral symmetry generated by
\begin{align}
\mathcal{C} = (-1)^{L/2} \prod_{i} e^{i\pi S^z_{2i+1}} = \prod_{i} e^{i\pi S^x_{i}} .
\label{eq:chiral}
\end{align}

From the first equality above, the chiral index can be interpreted as the parity of the count of $|z\rangle$ states on the odd sites of a given configuration.
The equivalence of the two forms follows from three facts. First, the first form commutes with every bond operator, $[\mathcal{C}, \mathcal{B}_j] = 0$ for all $j$. Second, multiplying it by the product of odd-bond operators $\prod_{j\,\in\,\mathrm{odd}} \mathcal{B}_j$ gives the second form. Third, this step uses the properties of the parity matrices, i.e., $\mathcal{P}^{z}_{j}\mathcal{P}^{y}_{j} = \mathcal{P}^{x}_{j}$ (see Appendix~\ref{app:Directorbasis}), which yields the second form. 

Its action reverses the transverse spin components on odd sites,
\begin{align}
\mathcal{C}S^x_{2i+1}\mathcal{C}^{-1}&=-S^x_{2i+1},&
\mathcal{C}S^y_{2i+1}\mathcal{C}^{-1}&=-S^y_{2i+1},
\end{align}
where $C=(-1)^{L/2} \prod_{i} e^{i\pi S^z_{2i+1}}$ and hence
\begin{equation}
\{\mathcal{C},H\}=0.
\end{equation}

The spectrum is therefore symmetric about zero energy. We focus on even-length chains with PBC throughout this section, except where stated otherwise; we discuss open boundary conditions (OBC) and odd-length periodic chains at the end.
Using a transfer-matrix formalism, we determine a lower bound on the number of zero modes in each bond-charge sector and the scaling of the total number summed over all bond-charge configurations. This bound matches exactly the numerically obtained zero modes after removing all the residual space-group symmetries, by introducing random bond disorder in the coupling strengths.

\subsection{Transfer Matrix formalism}

For each bond $(j,j+1)$, we define a transfer matrix whose entries specify
the allowed neighboring states:
\begin{equation}
T^{(w_j)}_{ki} =
\begin{cases}
1, & \begin{array}{@{}l@{}}
     \text{if states $k$ and $i$ are allowed}\\
     \text{on a bond with charge $w_j$},
     \end{array}\\
0, & \text{otherwise}.
\end{cases}
\end{equation}
For a periodic chain with bond configuration $\vec{w}=(w_1,\ldots,w_L)$, where
$w_j=\pm 1$, the dimension of the corresponding fragment is
\begin{equation}
D(w)=\operatorname{Tr}\!\left[\prod_{j=1}^{L}T^{(w_j)}\right].
\end{equation}
In the ordered Cartesian basis $(|x\rangle,|y\rangle,|z\rangle)$, the two transfer matrices are
\begin{equation}
T^{(+)} =
\begin{pmatrix}
0 & 1 & 1\\
1 & 0 & 0\\
0 & 1 & 1
\end{pmatrix},
\qquad
T^{(-)} =
\begin{pmatrix}
1 & 0 & 0\\
0 & 1 & 1\\
1 & 0 & 0
\end{pmatrix}.
\end{equation}
The row and column indices $1$, $2$, and $3$ correspond to the $x$, $y$, and $z$ states, respectively.
The Hamiltonian connects computational-basis states with opposite eigenvalues of the chiral operator $\mathcal{C}$ (see Appendix~\ref{app:Directorbasis} for details). Therefore, the Hilbert-space graph is bipartite, with its two sets comprising the $\mathcal{C}=+1$ and $\mathcal{C}=-1$ states. Ordering the basis by parity of $\mathcal{C}$ gives
\begin{equation}
H =
\begin{pmatrix}
0 & H_{AB}\\
H_{AB}^{\dagger} & 0
\end{pmatrix},
\end{equation}
where $H_{AB}$ is an $n_A\times n_B$ matrix, and $n_A$, $n_B$ are
the dimensions of the $\mathcal{C}=+1$ and $\mathcal{C}=-1$ subspaces,
respectively.

The bipartite structure guarantees zero modes whenever the two chiral subspaces have unequal dimensions. By the rank-nullity theorem,
\begin{equation}\label{eq:rank_nullity}
Z = \dim\ker H \geq |n_A-n_B|.
\end{equation}

To evaluate this imbalance in a bond-charge sector $\vec{w}$, we introduce the diagonal matrix,
\begin{equation}
D=
\begin{pmatrix}
1 & 0 & 0\\
0 & 1 & 0\\
0 & 0 & -1
\end{pmatrix},
\end{equation}
which assigns a chiral-parity weight to each local configuration. The signed imbalance is then obtained from the weighted transfer-matrix
product,
\begin{equation}\label{eq:imbalance_eqn}
\Delta_{\vec{w}}
=
\operatorname{Tr}\!\left[
T^{(w_1)}D\,T^{(w_2)}
\cdots
T^{(w_{L-1})}D\,T^{(w_{L})}
\right],
\end{equation}
and the resulting lower bound on the number of zero modes in this sector is given by the chiral index $\mathcal{N}_L(\vec{w})=|\Delta_{\vec{w}}|$.

To numerically count the zero modes enforced by sublattice imbalance in each bond-charge sector, we introduce weak bond disorder that breaks the residual spatial symmetries while preserving the chiral symmetry. Hence, the couplings are taken as
\begin{align}
\nonumber J_j = 1 + \delta_j, \qquad \delta_j \sim \mathrm{Uniform}(-W, W), \qquad W = 0.1
\end{align}
where $\delta_j$ are drawn independently. 

\begin{figure}
    \centering
    \includegraphics[width=\linewidth]{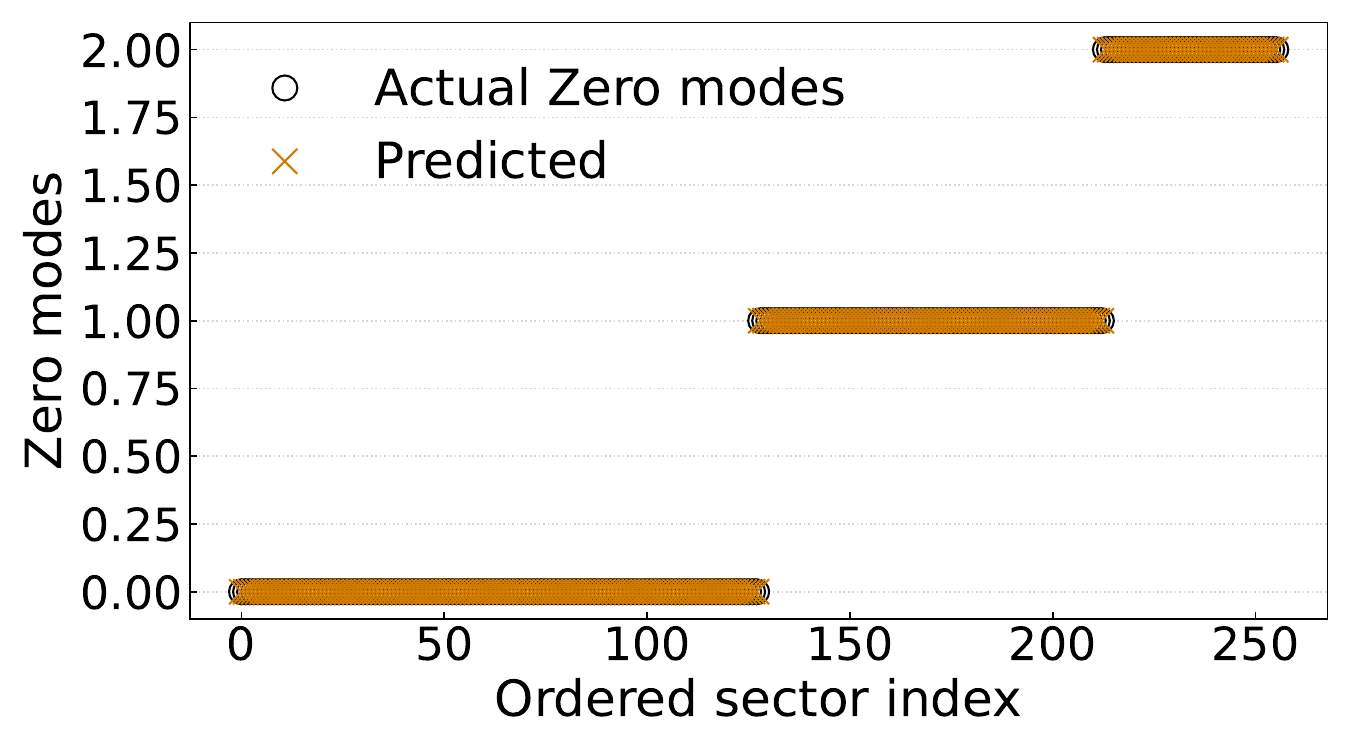}
    \caption{Zero-mode count in each bond-charge sector for $L=8$.
    The numerical count agrees with the predicted number given by chiral-imbalance bound
    $|\Delta_{\vec{w}}|$.}
    \label{fig:L8_sector_zm_scan}
\end{figure}

This serves the purpose of removing the need to resolve each sector further into spatial-symmetry subspaces, whose structure varies from sector to sector, which also lifts additional zero modes that are protected by these spatial symmetries of the clean chain, such as the antipodally correlated states of Sec.~\ref{sec:VolumeLawEntangledScars}, which require the two halves of the chain to carry identical couplings. The disorder is weak enough that all couplings remain positive and close to their initial value. For all system sizes examined, the number of exact zero modes in every sector then saturates the lower bound $|\Delta_{\vec{w}}| = |n_{A}-n_{B}|$ for a sector $\vec{w}$ given in Eq.~\eqref{eq:rank_nullity}, as illustrated for $L = 8$ in Fig.~\ref{fig:L8_sector_zm_scan}. We further find that the sector-resolved chiral index magnitude is at most two, irrespective of system size. 

\subsection{Full counting}

\begin{figure}
    \centering
    \includegraphics[width=0.94\linewidth]{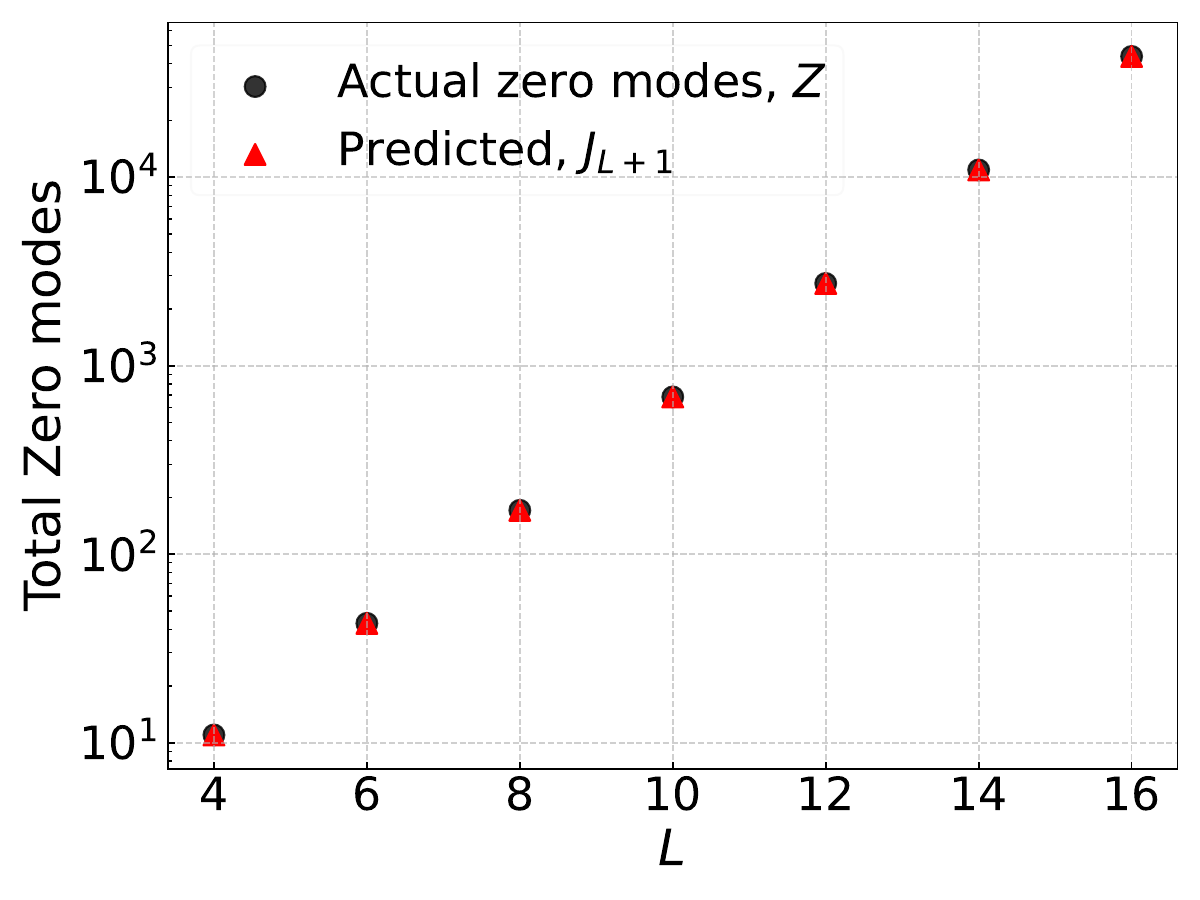}
    \caption{Sum of zero modes across all sectors obtained numerically $Z$  for the disordered Hamiltonian vs $L$. The predicted numbers are given by $J_{L+1}$ in Eq.~\eqref{eq:zm_count_jacobsthal}.}
    \label{fig:zm_count_proof}
\end{figure}

\begin{figure}
    \centering
    \includegraphics[width=0.9\linewidth]{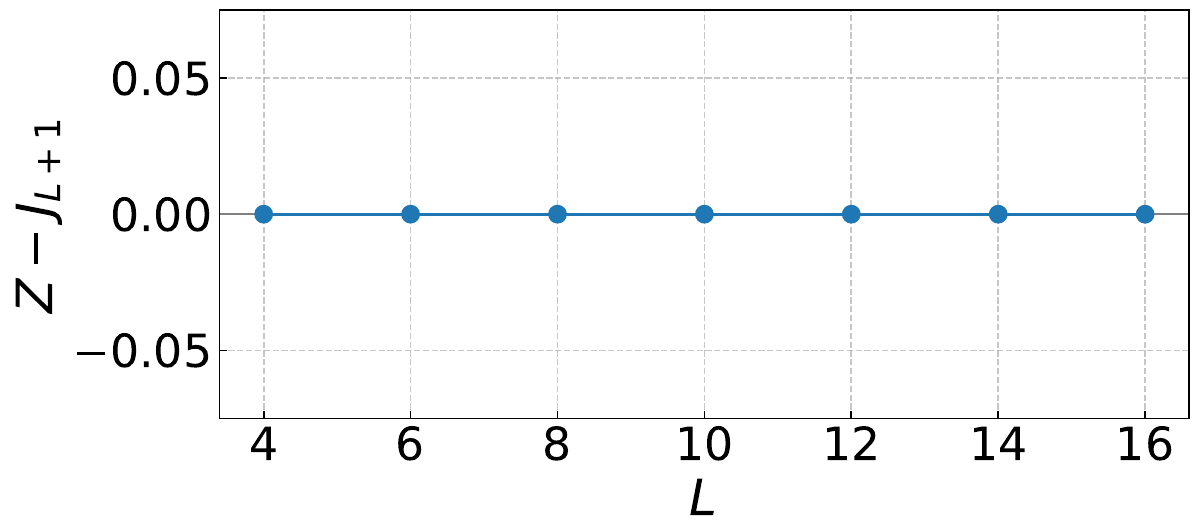}
    \caption{Difference between the number of zero modes obtained numerically $Z$ and the predicted count $J_{L+1}$ given by the Jacobsthal number (see App.~\ref{app:full_zm_count}).}
    \label{fig:zm_diff_scaling}
\end{figure}

Since the chiral-imbalance bound is saturated in every bond-charge sector, the total chiral index is the sum of the absolute values over all sectors, as shown below. 
\begin{equation}\label{eq:absolute_sum_zm}
\mathcal{N}_L=\sum_{\vec{w}}|\Delta_{\vec{w}}|.
\end{equation}
The absolute value prevents a direct evaluation by summing the transfer matrices over bond charges. Instead, we construct an alternative set of transfer matrices that closes under multiplication and reduces the sum to a six-state finite-state machine (FSM)~\cite{FSM_book}. As derived in Appendix~\ref{app:full_zm_count}, for even $L$ this yields
\begin{equation}\label{eq:zm_count_jacobsthal}
\mathcal{N}_L=\frac{2^{L+1}+1}{3} \equiv J_{L+1}
\end{equation}
This expression is the Jacobsthal number
$J_{L+1}$~\cite{weisstein_jacobsthal}. This implies the total chiral index $\mathcal{N}_{L}$ lower bounds the number of the zero modes, i.e. $Z\geq \mathcal{N}_{L}$. This lower bound using Eq.~\eqref{eq:zm_count_jacobsthal} agrees with
the numerically found chiral index for the disordered model as shown in Fig.~\ref{fig:zm_count_proof}, and the difference between the numerically obtained count $Z$ and predicted ($J_{L+1}$) is shown in Fig.~\ref{fig:zm_diff_scaling}, which is found to be $0$ for all system sizes probed, strengthening the understanding that the actual count from the chiral symmetry alone is $J_{L+1}$. 

\begin{figure}
    \centering
    \includegraphics[width=\linewidth]{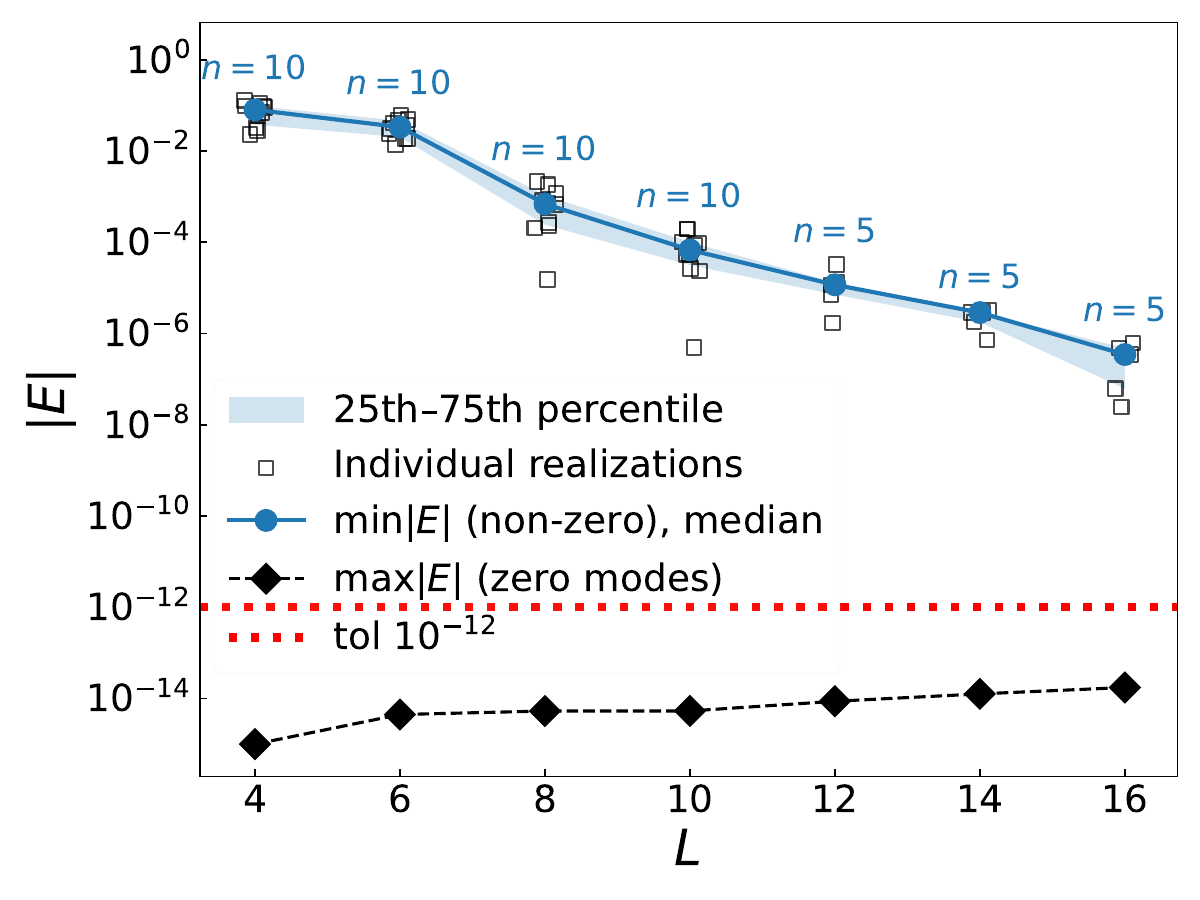}
    \caption{Separation of the numerical zero modes from the rest of the spectrum as a function of system size $L$. Open white squares show, for individual disorder realizations across all sectors, the smallest $|E|$ among the states not counted as zero modes; the solid line and shaded band give the median and the 25th--75th percentile range over realizations, and the number $n$ of realizations is indicated above the data at each $L$. Diamonds show the largest $|E|$ among the states counted as zero modes, maximized over realizations, which remains at machine precision. The red dotted line marks the tolerance $\mathrm{tol} = 10^{-12}$ used to identify the zero modes.}
    \label{fig:zm_gap_scaling}
\end{figure}

To verify that the zero-mode count does not depend on the numerical tolerance $\mathrm{tol} = 10^{-12}$, we track two quantities as functions of system size over several disorder realizations. The first is the largest $|E|$ among the states classified as zero modes. The second is the smallest $|E|$ among the remaining states, which measures how close the rest of the spectrum comes to $E = 0$. As shown in Fig.~\ref{fig:zm_gap_scaling}, the numerical zero modes remain at machine precision, $|E| \lesssim 10^{-14}$, for all system sizes, while the smallest nonzero $|E|$ stays several orders of magnitude above the tolerance for every realization. The two sets of eigenvalues are therefore cleanly separated, and the zero-mode count is insensitive to the precise choice of tolerance as a function of $L$.

The counting procedure also extends to open boundary conditions: the trace in the transfer-matrix expressions is replaced by appropriate boundary vectors, while the bulk transfer matrices and their algebra remain unchanged. Odd-length periodic chains require a modified definition of the chiral operator because the alternating bond pattern cannot close around the ring without two adjacent bonds of the same type. The subsequent counting follows the same construction, with an additional terminal ``letter" in the FSM, as discussed in Appendix~\ref{app:full_zm_count}.

\section{Antipodally correlated zero modes}
\label{sec:VolumeLawEntangledScars}

In this section, we give the exact analytic expression of $2^{L/2}$ exact zero-energy eigenstates. These states exist in sectors whose bond configuration is invariant under translation by $L/2$, and they follow from the chiral symmetry of the model through a construction analogous to those developed for the PXP model~\cite{Volume_law_scar_PXP} and other spin chains~\cite{Exact_volume_law_zero_modes}. Previous constructions have not addressed how these zero modes are distributed across symmetry sectors. Our formulation makes this distribution explicit: it shows which sectors host such states and why the construction does not apply in the others.

\subsection{The antipodally correlated states}

We write $L = 2N$ and label the sites of the second half of the chain by $\bar{i} \equiv i + N$, so that each site $i$ of the first half is paired with its antipode $\bar{i}$. Consider a bond configuration $\vec{w}$ of the full chain that is invariant under translation by $N$, i.e., $w_{\bar{j}} = w_j$. Such a configuration is fully specified by its restriction to the $N$ bonds of a periodic half chain. We denote by $\mathcal{F}_{\vec{w}}$ the set of Cartesian product states $|f\rangle = |f(1) f(2) \cdots f(N)\rangle$, with $f(j) \in \{x, y, z\}$, of the periodic half chain that belong to this half-chain bond configuration, and by $D_f = |\mathcal{F}_{\vec{w}}|$ their number.

On a Cartesian product state, the chiral operator acts as a sign. For the construction below it is convenient to use the second form of Eq.~\eqref{eq:chiral}, $\mathcal{C}_x = \prod_{j} e^{i\pi S^x_j}$, restricted to the half chain. Since $e^{i\pi S^x} = \mathrm{diag}(+1,-1,-1)$ in the Cartesian basis $(x, y, z)$,
\begin{align}
    \langle f | \mathcal{C}_x | f \rangle = (-1)^{N_{\bar{x}}(f)},
    \qquad
    N_{\bar{x}}(f) = \#\bigl\{ j : f(j) \neq x \bigr\},
\end{align}
where $N_{\bar{x}}(f)$ counts the sites not occupied by the state $|x\rangle$. The antipodally correlated zero mode in sector $\vec{w}$ is
\begin{align}\label{eq:Volumelawstate}
    |\psi\rangle_V
    = \frac{1}{\sqrt{D_f}}
    \sum_{f \in \mathcal{F}_{\vec{w}}}
    (-1)^{N_{\bar{x}}(f)}\,
    |f\rangle \otimes |f\rangle .
\end{align}
For $L \equiv 0 \pmod 4$, i.e., an even half-chain length $N$, the sign can equivalently be written as $(-1)^{N_z^{\mathrm{odd}}(f)}$, with $N_z^{\mathrm{odd}}(f)$ the number of odd sites occupied by $|z\rangle$, up to a sector-dependent global sign (see below).

\paragraph*{Entanglement.}
Any contiguous block of $N$ sites contains exactly one site from each antipodal pair. For such a bipartition, the reduced density matrix is maximally mixed on $\mathcal{F}_{\vec{w}}$, $\rho = \mathbb{1}_{D_f}/D_f$, and the entanglement entropy attains its maximal value within the sector, $S = \ln D_f$~\cite{Volume_law_scar_PXP, Exact_volume_law_zero_modes}. In sectors where $D_f$ grows exponentially with $N$, this constitutes a volume law. This is not true of every sector: in the sector with all $w_j = -1$, for example, $\mathcal{F}_{\vec{w}}$ contains only $|xx\cdots x\rangle$ and $|yy\cdots y\rangle$, so $|\psi\rangle_V$ reduces to a cat state with $S = \ln 2$. For a smaller contiguous subsystem $A$ within one half, $\rho_A$ is diagonal in the Cartesian basis, and its entropy equals the Shannon entropy of the configurations in $\mathcal{F}_{\vec{w}}$ restricted to $A$.

The maximal entanglement thus relies on each antipodal pair being split by the bipartition. A subsystem containing complete pairs, such as $A = \{1, \bar{1}, 2, \bar{2}\}$, behaves very differently. Because the two sites of each pair are perfectly correlated in $|\psi\rangle_V$, such a subsystem is entangled with the rest of the chain only through the bond constraints at the edges of the block, and its entanglement entropy remains of order one rather than growing with $N$. The volume-law entanglement of these states is therefore a consequence of their antipodal pairing structure, not a generic property of all bipartitions.

\paragraph*{Relation to ETH.}
For contiguous bipartitions, this entanglement is consistent with the eigenstate thermalization hypothesis (ETH). Because of the chiral symmetry, the spectrum is symmetric about zero, so $E = 0$ corresponds to infinite temperature. Since the Hilbert space is fragmented, the relevant form of ETH is Krylov-restricted thermalization~\cite{Sanjay_krylov_restricted_thermalization}, which predicts that the reduced density matrix of an eigenstate at this energy approaches the infinite-temperature state of its Krylov subspace. To leading order, this agrees with $\rho = \mathbb{1}_{D_f}/D_f$.

The antipodally correlated states nevertheless violate ETH for non-contiguous bipartitions. For a small subsystem such as $A = \{1, \bar{1}, 2, \bar{2}\}$, ETH predicts a reduced density matrix close to the infinite-temperature state, with entropy proportional to $|A|$. For $|\psi\rangle_V$, in contrast, $\rho_A$ is far from maximally mixed and carries only $O(1)$ entanglement with the rest of the chain. Equivalently, since $|\psi\rangle_V$ is a superposition of configurations with $f(i) = f(\bar{i})$, antipodal sites remain perfectly correlated despite being separated by a distance $L/2$, whereas connected correlations in a thermal state are expected to decay exponentially with distance. The volume-law entanglement of these states thus coexists with a clear violation of ETH, and both originate from the same antipodal pairing structure.

\begin{figure}
    \centering
    \includegraphics[width=0.9\linewidth]{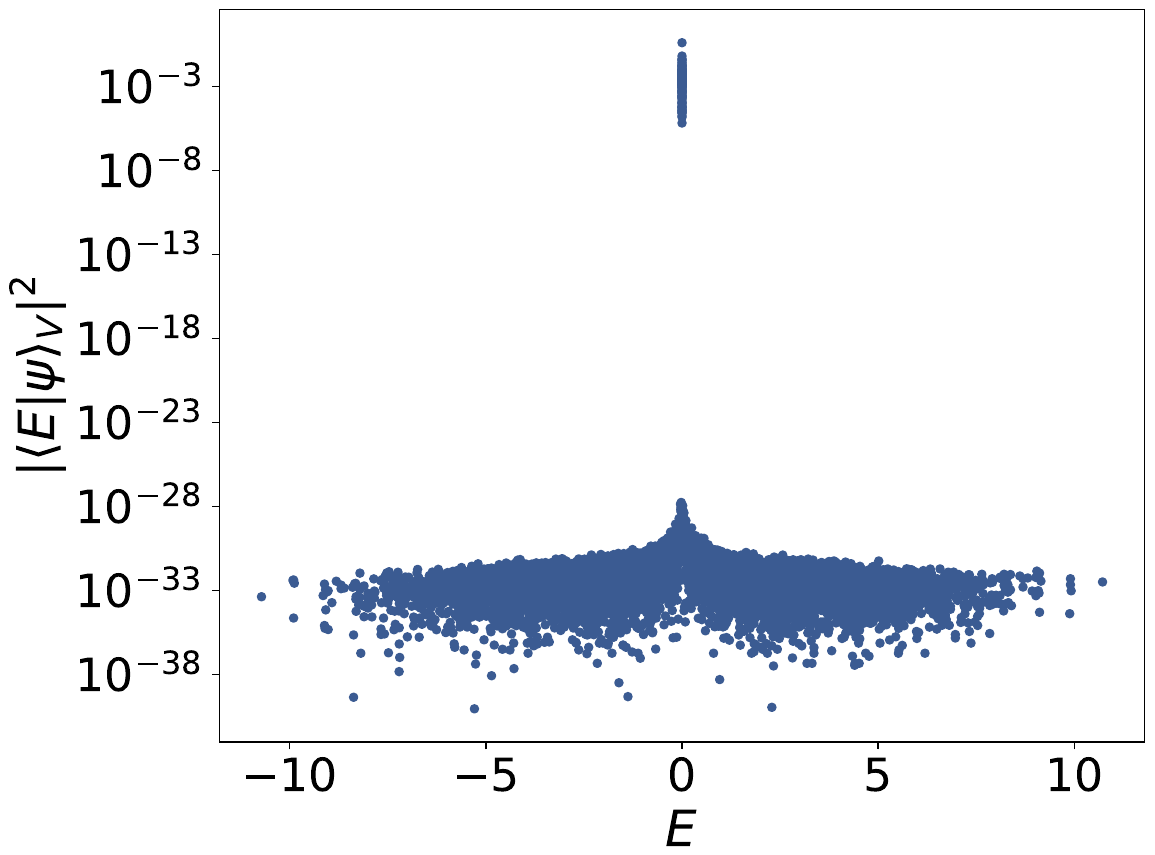}
    \caption{Overlap of the state in Eq.~\eqref{eq:Volumelawstate} with the eigenstates of the sector with bond motif $(1,1,1,-1,-1,-1)$ for $L = 24$ (sector dimension $10402$). All weight lies at $E = 0$.}
    \label{fig:volume_law_proof}
\end{figure}

\paragraph*{Numerical check and counting.}
These states also exist in sectors that do not host the approximate scars of Sec.~\ref{sec:SectorResolvedApproximateScarStudy}. An example is the sector of the $L = 24$ chain with the repeating bond motif $(1, 1, 1, -1, -1, -1)$, defined in Sec.~\ref{sec:SectorResolvedApproximateScarStudy}. As shown in Fig.~\ref{fig:volume_law_proof}, the overlap of the state in Eq.~\eqref{eq:Volumelawstate} with the eigenstates of this sector is supported entirely on the zero-energy manifold, which provides an independent numerical check of the construction. Because antipodal bonds must carry the same charge, while the $N$ bond charges of the half chain can be chosen freely, the construction yields one such state for each of the $2^N$ half-chain bond configurations whose set $\mathcal{F}_{\vec{w}}$ is nonempty.

In the remainder of this section, we prove that $|\psi\rangle_V$ is an exact zero mode using the operator--state mapping~\cite{Jamiolkowski1972, Choi1975, Preskill1998-aq}. Under this mapping, the chiral operator of a half chain becomes a state of the full chain, and $|\psi\rangle_V$ is precisely this state projected onto the sector $\vec{w}$.

\subsection{Operator--state mapping}

We briefly review the operator--state mapping, following Ref.~\cite{Preskill1998-aq}. Fix an orthonormal basis $\{|a\rangle\}$ of a Hilbert space $\mathcal{H}$; throughout, we take this to be the Cartesian basis of Eq.~\eqref{eq:Directorbasis}, and all transposes below are taken in it. An operator is mapped to a vector in the doubled space $\mathcal{H} \otimes \mathcal{H}$ by
\begin{align}
    \hat{O} = \sum_{a,b} O_{ab}\, |a\rangle\langle b|
    \quad\longmapsto\quad
    |O\rangle\!\rangle = \sum_{a,b} O_{ab}\, |a\rangle \otimes |b\rangle .
\end{align}
Strictly, the second factor is the dual space $\mathcal{H}^*$. We identify it with a second copy of $\mathcal{H}$ through the Cartesian basis, $\langle b| \mapsto |b\rangle$. This identification is a choice, and the basis enters the construction only through the transposes below.

We also highlight an identity: i.e., left and right multiplication of an operator become actions on the first and second factors.
\begin{align}\label{eq:vectorization_identity}
    (A \otimes B)\, |O\rangle\!\rangle = |A O B^{T}\rangle\!\rangle .
\end{align}
In particular, for any operator $H$ on $\mathcal{H}$,
\begin{align}\label{eq:choi_H_action}
    \big( H \otimes \mathbb{1} + \mathbb{1} \otimes H \big)\, |O\rangle\!\rangle
    = |H O + O H^{T}\rangle\!\rangle .
\end{align}
For a single site, the vectorized identity $|\mathbb{1}\rangle\!\rangle = \sum_a |a\rangle \otimes |a\rangle$ obeys the transfer identity
\begin{align}\label{eq:transfer_identity}
    (A \otimes B)\, |\mathbb{1}\rangle\!\rangle
    = (A B^{T} \otimes \mathbb{1})\, |\mathbb{1}\rangle\!\rangle ,
\end{align}
which moves an operator from the dual space to the Hilbert space or vice versa at the cost of a transpose.

In the Cartesian basis, the spin-$1$ matrices are purely imaginary and antisymmetric, $(S^a)_{bc} = -i\epsilon_{abc}$, so $(S^a)^T = -S^a$. Consequently, any Hamiltonian built from spin bilinears, including Eq.~\eqref{eq:TranslationalinvariantKitaev}, satisfies $H^T = H$, while $e^{i\pi S^a}$ is real and diagonal. As we show below, the construction holds in any product basis, but the Cartesian basis is the one in which its hypotheses reduce to the ordinary chiral symmetry $\{\mathcal{C}, H\} = 0$.

\subsection{Construction and proof}

We consider a periodic chain of $2N$ sites, with identical local Hilbert spaces $\mathcal{H}_i \cong \mathcal{H}_{\bar{i}} \cong \mathbb{C}^{d_i}$, and a nearest-neighbor Hamiltonian
\begin{align}\label{eq:physical_H}
    H(2N)
    = \sum_{i=1}^{N-1} h_{i,i+1}
    + h_{N,\bar{1}}
    + \sum_{i=1}^{N-1} h_{\bar{i},\overline{i+1}}
    + h_{\bar{N},1} ,
\end{align}
following the notation of Refs.~\cite{Volume_law_scar_PXP, Exact_volume_law_zero_modes}. The two halves carry identical bulk bonds, $h_{\bar{i},\overline{i+1}} = h_{i,i+1}$, and are joined by the two \emph{seam} bonds $h_{N,\bar{1}}$ and $h_{\bar{N},1}$. Each bond is decomposed into products of single-site operators,
\begin{align}
    h_{i,j} = \sum_{\mu} A^{\mu}_i B^{\mu}_j ,
\end{align}
where, e.g., $\mu \in \{x, y, z\}$ for spin-component couplings. Finite-range interactions are treated identically after decomposing them into products of single-site operators.

Let $\mathcal{C} = \bigotimes_{i=1}^{N} C_i$ be an on-site unitary on the half chain, and define the local transfer map.
\begin{align}\label{eq:tau}
    \tau_i(X) C_{i} \equiv C_i X^{T}
\end{align}
Since the transpose of a tensor product of operators on distinct sites is the tensor product of their transposes,
\begin{align}\label{eq:tau_factorize}
    \mathcal{C} \Big( \prod_i X_i \Big)^{T}  = \prod_i \tau_i(X_i) \mathcal{C}.
\end{align}

We note that, for an onsite invertible symmetry, the transfer map is identical to the following:
\begin{align}
    \tau_i(X)  \equiv C_i X^{T} C^{-1}_{i}
\end{align}

\begin{theorem}\label{thm:choi}
Suppose that
\begin{enumerate}
\item[(i)] the local factors in the onsite chiral symmetry satisfy $C_i^{T} = \pm C_i$, with the transpose taken in the fixed product basis;
\item[(ii)] every bond of $H(2N)$, including both seam bonds, is chiral with respect to the transfer maps of the sites it connects, with seam sites $\bar{i}$ assigned the map $\tau_i$:
\begin{align}\label{eq:bulk_chiral}
    (\tau_i \otimes \tau_j)(h_{i,j}) \equiv \sum_\mu \tau_i(A^\mu)\, \tau_j(B^\mu) = -\, h_{i,j} .
\end{align}
\end{enumerate}
Then the normalized Choi state of $\mathcal{C}$,
\begin{align}\label{eq:choi_state}
    |C\rangle\!\rangle
    &= \frac{1}{\sqrt{D}}\, |\mathcal{C}\rangle\!\rangle
    = (\mathcal{C} \otimes \mathbb{1}) \bigotimes_{i=1}^{N} |\Lambda\rangle_{i\bar{i}}, \\
    \qquad
    |\Lambda\rangle_{i\bar{i}} &= \frac{1}{\sqrt{d_i}} \sum_{a=1}^{d_i} |a\rangle_i \otimes |a\rangle_{\bar{i}} ,
\end{align}
with $D = \prod_i d_i$, is an exact zero-energy eigenstate of $H(2N)$.
\end{theorem}

The idea of the proof is to show that, on $|C\rangle\!\rangle$, the physical Hamiltonian acts as two identical copies of an \emph{auxiliary} periodic half-chain Hamiltonian $\tilde{H}(N)$, one on each half (Steps 1 and 2). Chirality of $\tilde{H}(N)$ (Step 3) then makes $|C\rangle\!\rangle$ a zero mode through Eq.~\eqref{eq:choi_H_action} (Step 4).

\begin{proof}
\emph{Step 1: first seam.} The state $|C\rangle\!\rangle$ is a product of pairs $(i, \bar{i})$, and the seam bond $h_{N,\bar{1}}$ acts only on the pairs $(N, \bar{N})$ and $(1, \bar{1})$. Applying the transfer identity, Eq.~\eqref{eq:transfer_identity}, and the definition of the transfer map, Eq.~\eqref{eq:tau}, to the pair $(1, \bar{1})$,
\begin{align}
    \big( C_1 \otimes B^{\mu}_{\bar{1}} \big) |\Lambda\rangle_{1\bar{1}}
    = \big( C_1 (B^{\mu}_1)^{T} \otimes \mathbb{1} \big) |\Lambda\rangle_{1\bar{1}} \\
    \nonumber = \big( \tau_1(B^{\mu}_1)\, C_1 \otimes \mathbb{1} \big) |\Lambda\rangle_{1\bar{1}} ,
\end{align}
where $B^\mu_1$ denotes the copy of $B^\mu_{\bar{1}}$ acting on site $1$. Hence
\begin{align}\label{eq:push_through}
    h_{N,\bar{1}}\, |C\rangle\!\rangle = \tilde{h}_{N,1}\, |C\rangle\!\rangle ,
    \qquad
    \tilde{h}_{N,1} \equiv \sum_\mu A^{\mu}_N\, \tau_1(B^{\mu}_1) .
\end{align}
On $|C\rangle\!\rangle$, the seam bond therefore acts as an effective closing bond $\tilde{h}_{N,1}$ that lies entirely within the first half.

\emph{Step 2: second seam.} For $h_{\bar{N},1} = \sum_\mu A^\mu_{\bar{N}} B^\mu_1$, the factor on site $1$ must instead be transferred to site $\bar{1}$. Running the same steps in reverse gives
\begin{align}
    \big( B^\mu_1 \otimes \mathbb{1} \big) |C\rangle\!\rangle
    &= \big( \mathbb{1} \otimes \bar{\tau}_1(B^\mu_{\bar{1}}) \big) |C\rangle\!\rangle , \\
    \qquad
    \nonumber \bar{\tau}_i(X) &\equiv \big( C_i^{-1} X C_i \big)^{T} = C_i^{T} X^{T} \big( C_i^{T} \big)^{-1} .
\end{align}
By hypothesis~(i), $\bar{\tau}_i = \tau_i$, so $h_{\bar{N},1}$ acts on $|C\rangle\!\rangle$ as $\tilde{h}_{\bar{N},\bar{1}} = \sum_\mu A^\mu_{\bar{N}}\, \tau_1(B^\mu_{\bar{1}})$, a closing bond of the same form as $\tilde{h}_{N,1}$, now within the second half. Defining the auxiliary Hamiltonian on the periodic half chain,
\begin{align}\label{eq:single_replica_H}
    \tilde{H}(N) = \sum_{i=1}^{N-1} h_{i,i+1} + \tilde{h}_{N,1} ,
\end{align}
Steps 1 and 2 together give
\begin{align}\label{eq:boundary_matching}
    H(2N)\, |C\rangle\!\rangle = \big( \tilde{H}(N) \otimes \mathbb{1} + \mathbb{1} \otimes \tilde{H}(N) \big) |C\rangle\!\rangle .
\end{align}

\emph{Step 3: chirality of the auxiliary Hamiltonian.} By Eq.~\eqref{eq:tau_factorize}, the condition
\begin{align}\label{eq:chiral_condition}
    \tilde{H}(N)\, \mathcal{C} + \mathcal{C}\, \tilde{H}(N)^{T} = 0
\end{align}
is equivalent to $\big( \bigotimes_i \tau_i \big)\big(\tilde{H}(N)\big) = -\tilde{H}(N)$, which can be verified bond by bond. The bulk bonds satisfy it by hypothesis~(ii). For the closing bond, note first that hypothesis~(i) makes each $\tau_i$ an involution,
\begin{align}
    \tau_i^2(X) = C_i \big( C_i^{T} \big)^{-1} X\, C_i^{T} C_i^{-1} = X .
\end{align}
This implies that the convolution of the transfer map $\tau_{i}^{2} = \rm id$, where $\rm id$ is an identity map, and not to be confused with $\mathbb{1}$, which is an operator identity living on $\mathcal{H}$. Applying $\mathrm{id} \otimes \tau_1$ to the chirality of the seam bond, $\sum_\mu \tau_N(A^\mu)\, \tau_1(B^\mu) = -\sum_\mu A^\mu B^\mu$, and using $\tau_1^2 = \mathrm{id}$ gives
\begin{align}
    \sum_\mu \tau_N(A^\mu)\, B^\mu = -\sum_\mu A^\mu\, \tau_1(B^\mu) = -\tilde{h}_{N,1} .
\end{align}
The left-hand side equals $(\tau_N \otimes \tau_1)(\tilde{h}_{N,1})$, so the closing bond is chiral and Eq.~\eqref{eq:chiral_condition} holds.

\emph{Step 4: conclusion.} Combining Eqs.~\eqref{eq:boundary_matching}, \eqref{eq:choi_H_action}, and \eqref{eq:chiral_condition},
\begin{align}
    H(2N)\, |C\rangle\!\rangle
    = \frac{1}{\sqrt{D}}\, \big| \tilde{H}(N)\, \mathcal{C} + \mathcal{C}\, \tilde{H}(N)^{T} \big\rangle\!\big\rangle
    = 0 .
\end{align}
\end{proof}

\paragraph*{Application to the Kitaev chain.}
For the Hamiltonian in Eq.~\eqref{eq:TranslationalinvariantKitaev}, $H = \sum_j S^x_j S^y_{j+1}$, we take the half-chain chiral operator $\mathcal{C}_x = \prod_{j=1}^{N} e^{i\pi S^x_j}$, so that $C_j = e^{i\pi S^x_j}$ on every site. Each $C_j$ is real and diagonal in the Cartesian basis, so hypothesis~(i) holds with $C_j^T = C_j$. Using $(S^a)^T = -S^a$ together with Eq.~\eqref{eq:parity_relations}, the transfer map acts as
\begin{align}
    \tau_j(S^x) = -S^x, \qquad \tau_j(S^y) = S^y, \qquad \tau_j(S^z) = S^z,
\end{align}
on every site. Every bond, including the seam bonds viewed as the closing bond $(N, 1)$ of the half chain, therefore satisfies $(\tau_j \otimes \tau_{j+1})(S^x_j S^y_{j+1}) = -S^x_j S^y_{j+1}$, and hypothesis~(ii) holds for any $N$. Moreover, $\tau_1(S^y) = S^y$, the closing bond $\tilde{h}_{N,1} = S^x_N S^y_1$ coincides with a bulk bond and $\tilde{H}(N)$ is simply the periodic $N$-site chain. Since $H^T = H$ in the Cartesian basis, the chirality condition, Eq.~\eqref{eq:tau}, reduces to $\{\mathcal{C}_x, \tilde{H}(N)\} = 0$, the ordinary chiral symmetry of the half chain.

The sublattice form $\mathcal{C}_z = \prod_{j\,\in\,\mathrm{odd}} e^{i\pi S^z_j}$ used elsewhere in this work satisfies hypothesis~(ii) only when $N$ is even. For odd $N$, the closing bond connects two odd sites, on which $\tau_j(S^{x,y}) = S^{x,y}$, so that $(\tau_N \otimes \tau_1)(S^x_N S^y_1) = +S^x_N S^y_1$, and the construction fails. For even $N$, using $P^x = P^y P^z$, the two forms on the half chain are related by $\mathcal{C}_x = \mathcal{C}_z \prod_{j\,\in\,\mathrm{odd}} \mathcal{B}_j$. Within a sector, the product of bond operators is the constant $\prod_{j\,\in\,\mathrm{odd}} w_j = \pm 1$, so both forms give the same state up to a global sign.

\paragraph*{Sector resolution.}
The bond operators $\mathcal{B}_j$ are real and diagonal in the Cartesian basis and commute with both $\mathcal{C}$ and $H(2N)$. Projecting $|C\rangle\!\rangle$ onto a bond sector $\vec{w}$ of the full chain therefore yields another zero mode. By Eq.~\eqref{eq:vectorization_identity}, a bond operator on the first half can be moved to the antipodal bond without change, $(\mathcal{B}_j \otimes \mathbb{1}) |C\rangle\!\rangle = (\mathbb{1} \otimes \mathcal{B}_{\bar{j}}) |C\rangle\!\rangle$, and the same holds for the two seam bonds and the closing bond of $\tilde{H}(N)$. The projection is thus nonzero only if $w_{\bar{j}} = w_j$ for all $j$, i.e., only in sectors invariant under translation by $L/2$. In such a sector, since $\mathcal{C}$ is diagonal in the Cartesian basis with $\langle f | \mathcal{C} | f \rangle = \pm (-1)^{N_z^{\mathrm{odd}}(f)}$, the normalized projection is exactly the state in Eq.~\eqref{eq:Volumelawstate}. States in different sectors are orthogonal, so the construction yields $2^{L/2}$ linearly independent zero modes, one for each half-chain bond configuration with $D_f > 0$. Sectors that are not invariant under translation by $L/2$ may host other zero modes, but not ones of this form.

\section{Approximate quantum Scars in different sectors}
\label{sec:SectorResolvedApproximateScarStudy}

Throughout this and the next section, we refer to the $|z\rangle$ state as a \emph{defect}, while the $|x\rangle$ and $|y\rangle$ states are regarded as the \emph{vacua}. The motivation for this terminology is that, in the sector $\vec{w}=\{-1,-1,\dots,-1\}$, the product states $|xxx\dots\rangle$ ,$|yyy\dots\rangle$ are the only defect-free states. Throughout this work we refer to a local bond charge with $w_{j}=+1$ a \emph{wall} because a defect ($|z\rangle$ state) cannot move across a wall under the action of the Hamiltonian in Eq.~\eqref{eq:TranslationalinvariantKitaev}. States in different sectors can therefore be viewed as arising from the creation, annihilation, and hopping of defects, i.e., $|z\rangle$ states on top of these vacua. 

It has been established that, in the sector $\vec{w} = \{1, 1, 1, \dots\ ,1\}$, the model maps onto the PXP model~\cite{You_scars_in_Kitaev_model} with an effective spin-$1/2$ description~\cite{Spin-1_Kitaev_model_Dhar}. Consequently, this sector hosts QMBS, as illustrated in Ref.~\cite{You_scars_in_Kitaev_model}. QMBS have also been identified in other sectors~\cite{Balram_Kitaev_scars}; for instance, the sector $\vec{w} = \{1, 1, -1, \dots, 1, 1, -1\}$.
    
To begin the analysis, we start by systematically constructing product states within each sector that have a large overlap with the scarred eigenstates of that sector and hence display fidelity revivals. We refer to these states as the sector-Neel states, denoted by $|N\rangle_{\vec{w}}$, where $\vec{w}$ specifies the bond configuration. This construction is applicable only to bond configurations containing an even number of walls. We further illustrate the construction with several examples and demonstrate that it correctly reproduces the sector-Neel states in sectors where quantum many-body scars have previously been identified~\cite{Balram_Kitaev_scars, You_scars_in_Kitaev_model}.

We show that our constructed initial states exhibit fidelity revivals in three distinct classes of sectors, extending the known examples to a previously unexplored class of sectors. Furthermore, we demonstrate that the observed dynamics in all three classes can be understood within a universal formulation of the FSA~\cite{Turner_2018_PXP}.
For a quick summary of the findings of sectors studied in this section and Sec.~\ref{sec:NonScarredSectors}, the reader can refer to table.~\ref{tab:summarytable}.

\subsection{Initial state with support over the scar subspace}\label{sec:sectorneel}

For sectors containing an even number of walls, we construct the initial state as the completely defect-free state, i.e., a state containing no $|z\rangle$ state and composed entirely of the $|x\rangle$, $|y\rangle$ states. We refer to these states as the sector-Neel states, as in a lot of sectors these states show scarring dynamics, as discussed below. We present several examples to illustrate this construction.

Before illustrating the construction, a natural question arises, which is whether a given sector can admit more than two defect-free states. As we show in Appendix~\ref{app:proof}, this is not possible. Every sector containing an even number of walls possesses exactly two defect-free states, both of which can be constructed uniquely from the conserved bond operators.

We emphasize that although every sector containing an even number of walls possesses exactly two defect-free states, the existence of these states alone does not guarantee non-thermalizing dynamics. In general, most even-wall sectors do not exhibit fidelity revivals when initialized in a defect-free state. Instead, we find three distinct classes of bond configurations that support defect-free initial states that display robust fidelity revivals and the associated non-thermalizing behavior, which we discuss in the following one by one. We stress that these classes are not strict criteria for scarring: some sectors belonging to one of the classes nevertheless show no scarred dynamics when initialized in the sector-Neel state.

\begin{table*}
\caption{Summary of the sector classes studied in Sec.~\ref{sec:SectorResolvedApproximateScarStudy} and Sec.~\ref{sec:NonScarredSectors}. Bond motifs repeat periodically, with $w_j=+1$ denoting a wall. Sector-Neel states, defined in Sec.~\ref{sec:sectorneel}, can undergo fidelity revivals in certain classes of sectors listed below. They are given over one period of the bond motif, up to lattice translations, with $a^{n}$ denoting $n$ consecutive sites in the state $|a\rangle$. Sectors marked with $^\dagger$ were studied previously. Details of individual sectors are given in the indicated figures and appendices.}
\label{tab:sectors}
\centering
\renewcommand{\arraystretch}{1.2}
\begin{ruledtabular}
\begin{tabular}{p{0.19\linewidth}p{0.19\linewidth}p{0.15\linewidth}p{0.37\linewidth}}
Sector class & Bond motif & Sector-Neel state & Revivals from sector-Neel state \\
\hline
$k$-wall periodic & $\{1,(-1)^{k}\}$ & $|x^{k+1}y^{k+1}\rangle$ & Present for all $k = 0,\dots,5$ studied; quality varies with $k$ without a clear trend. $k=0$ (PXP)$^\dagger$~\cite{You_scars_in_Kitaev_model}, $k=1^\dagger$~\cite{Balram_Kitaev_scars}. [Fig.~\ref{fig:3wallfide}, App.~\ref{app:kwallnumerics}] \\
$k$-bound-wall & $\{1,1,(-1)^{k}\}$ & $|xy\,x^{k}\rangle$ & Present for $k = 1,\dots,6$; weaken with increasing $k$. $k=1,2^\dagger$~\cite{Balram_Kitaev_scars}. [Fig.~\ref{fig:2boundwallfide}, App.~\ref{app:boundwallnumerics}] \\
Mixed periodic
  & \raggedright Isolated and bound walls,\newline e.g.\ $\{1,1,-1,1,-1\}$
  & e.g.\ $|xy\,x^{2}y^{2}xy^{2}x\rangle$
  & \raggedright Present in most motifs studied; sensitive to the motif, absent for $\{1,1,(-1)^3,1,1,-1\}$. [Fig.~\ref{fig:Mixedsectorfide}, App.~\ref{app:mixedsectornumerics}]
  \tabularnewline
$\geq 3$ consecutive walls & e.g.\ $\{1,1,1,(-1)^{3}\}$ & e.g.\ $|xyxy^{4}xyx^{3}\rangle$ & Absent in all cases studied (at most isolated peaks). [Sec.~\ref{sec:NonScarredSectors}, App.~\ref{app:Nonscarred_Sectors}] \\
\end{tabular}
\end{ruledtabular}
\label{tab:summarytable}
\end{table*}

\begin{figure}
    \centering
    \includegraphics[width=\linewidth]{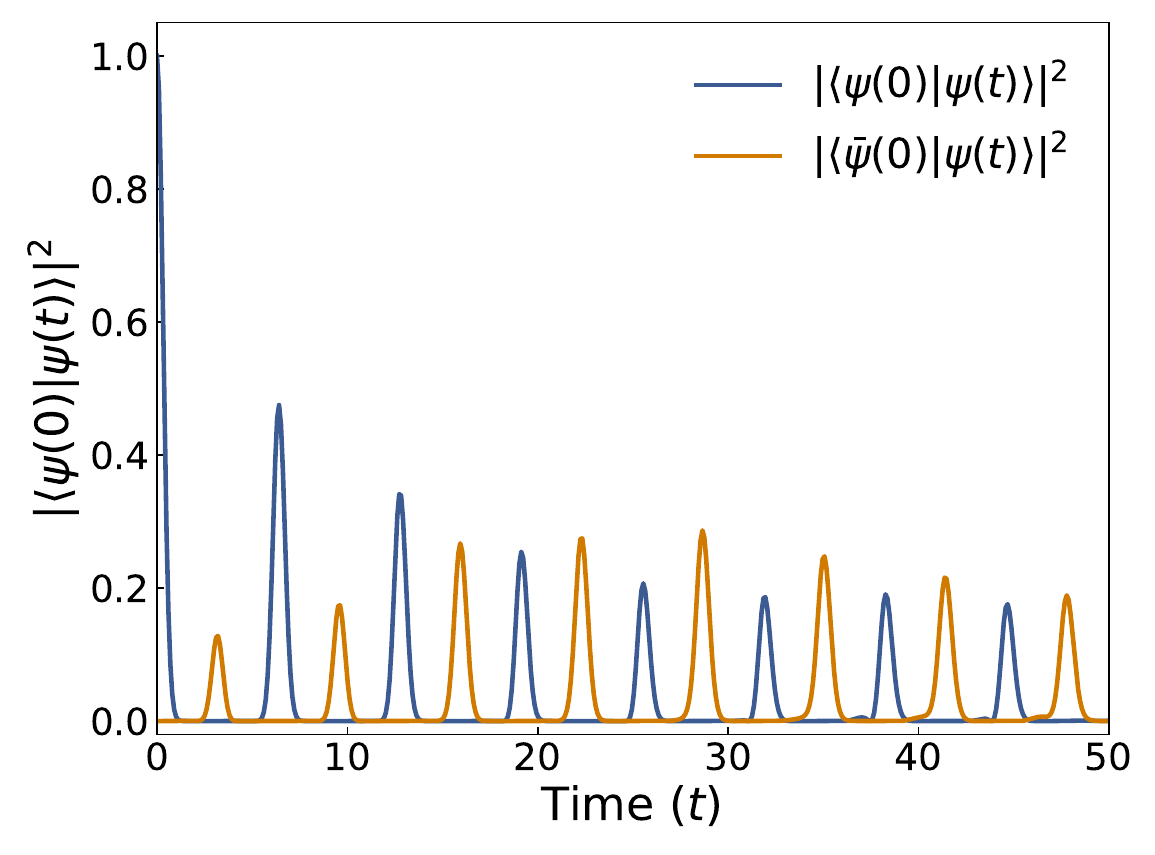}
    \caption{Fidelity plot for the mixed sector $w=\{1, -1, 1, -1, -1, \dots\}$, with L=25 (dim=30248 states). $|\psi\rangle$ represents the sector-Neel state, while $|\bar{\psi}\rangle$ represents the sector-Anti-Neel state (obtained by replacing $|x\rangle$ with $|y\rangle$)}
    \label{fig:Mixedsectorfide}
\end{figure}

\subsubsection{$k$-wall periodic sectors}\label{sec:kwallsector}

We define a $k$-wall periodic sector as a sector in which the walls, i.e. $w_j=+1$, appear periodically after every $k$ bonds, namely
$w_{k-wall}=\{1, \underbrace{-1, \dots}_{k}, 1, \underbrace{-1, \dots}_{k}, \dots,-1 \}$. Thus, the conserved bond charges repeat with a fixed periodic pattern along the chain. We refer to this repeating pattern of bond charges as the \emph{bond motif}. As discussed in the previous subsection, once the bond motif is specified, the corresponding sector-Neel state can be constructed uniquely using our prescription, which we denote as $|N\rangle_{\{w\}}$, where $\{w\}$ is the bond motif. We note that the $k$-wall periodic sectors exist only for system sizes commensurate with their bond motif. A motif of length $k+1$ requires $(k+1)$ divides $L$, and the existence of the sector-Neel state further requires an even number of walls, $L/(k+1) \in 2\mathbb{Z}$. Our numerical analysis is therefore restricted to such commensurate lengths, which limits the range of system sizes accessible for each $k$.

We further define the $0$-wall periodic sector as $w=\{+1,+1,\dots,+1\}$, which coincides with the PXP sector. Applying the construction given in Appendix~\ref{app:proof} to the PXP sector, the corresponding sector-Neel state is given by

\begin{align}
    |N\rangle_{\{1\}} \equiv |xy\rangle \equiv |xyxyxy\dots\rangle
\end{align}

As another example, we consider the $1$-wall periodic sector, characterized by the bond configuration $w=\{-1,1,-1,1,\dots,-1,1\}$. This sector has been studied in Ref.~\cite{Balram_Kitaev_scars}, where it was shown to host PXP-like approximate scar states. Applying the same construction, we find that the corresponding sector-Neel state is given by
\begin{align}
|N\rangle_{\{-1, 1\}} \equiv |xxyy\rangle \equiv |xxyyxxyy\dots\rangle
\end{align}

\begin{figure}[t]
    \centering

    \begin{subfigure}{0.48\linewidth}
        \centering
        \includegraphics[width=0.7\linewidth]{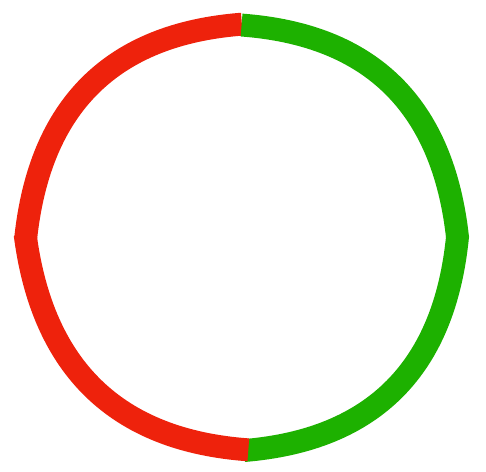}
        \caption{}
        \label{fig:2wallneel}
    \end{subfigure}
    \hfill
    \begin{subfigure}{0.48\linewidth}
        \centering
        \includegraphics[width=0.7\linewidth]{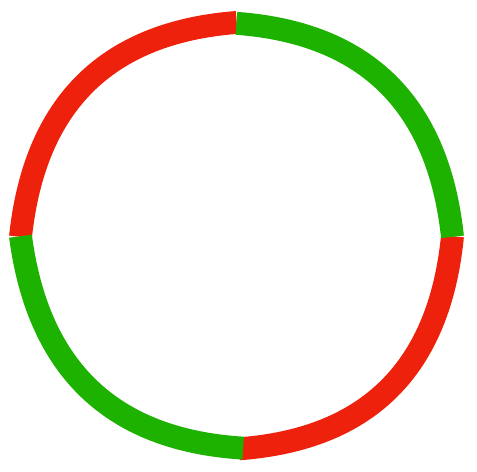}
        \caption{}
        \label{fig:4wall}
    \end{subfigure}

    \caption{Representative examples of sector-Neel states. (a) shows a state with two antipodal walls or $L/2$ wall periodic sector, while (b) shows the state with 4 walls having arcs of equal length or $L/4$ wall periodic sector. Green corresponds to the $|y\rangle$ states, and red corresponds to $|x\rangle$ states. Walls are located at the interfaces of the two colours.}
    \label{fig:TwoAndFourWallVaccumStates}
\end{figure}

In this work, we generalize this analysis to the remaining $k$-wall periodic sectors with $k=2,3,\dots,L/4,L/2$. We show that these sectors likewise exhibit a similar scar phenomenology. Representative sector-Neel states in the $L/4$-wall periodic and $L/2$-wall periodic sectors are illustrated in Fig.~\ref{fig:TwoAndFourWallVaccumStates}.

\subsubsection{k-bound wall periodic sectors}\label{sec:kboundwallsector}

There are also several sectors that do not fall into the above class but nevertheless exhibit PXP-like scar phenomenology. One such example is the sector with the bond-charge configuration $w=\{1,1,-1,\dots,1,1,-1\}$, which was studied in Ref.~\cite{Balram_Kitaev_scars}. This sector also hosts scar states originating from the sector—Neel state. Using the same construction, the corresponding sector-Neel state is given by 
\begin{align}
|N\rangle_{\{1, 1, -1\}} \equiv |xyx\rangle \equiv |xyxxyx\dots\rangle
\end{align}

More generally, we find that a broader family of sectors with bond configurations of the form $w=\{1, 1, \underbrace{-1, \dots}_{k}, 1, 1, \underbrace{-1, \dots}_{k}, \dots ,-1\}$ also exhibits PXP-like fidelity revivals. These correspond to periodic sectors in which pairs of adjacent walls, i.e. $1,1$, are separated by $k$ consecutive no-wall bonds, i.e. $w=-1$. Despite their distinct bond motifs, these sectors are described by essentially the same forward-scattering approximation discussed in Sec.~\eqref{sec:FSA}, and the construction of the corresponding sector-Neel states follows exactly the same prescription.

\subsubsection{Mixed periodic sectors}\label{sec:Mixedsector}

Somewhat surprisingly, we also find sectors with more intricate periodic structures, in which a bond motif containing both bound walls and single walls is repeated throughout the chain. More generally, these motifs may contain a combination of both or single wall or single bound wall, separated by arbitrary distances greater than or equal to 1, but are periodic. As an illustrative example, we consider the bond configuration $w= \{1, -1, 1, -1, -1, 1, -1, 1, -1, -1, \dots \}$, corresponding to the repeating bond motif $\{ 1, -1, 1, -1, -1\}$. Applying our construction, the corresponding sector-Neel state is
\begin{align}\label{eq:mixedsectorneel}
|N\rangle_{\{1, -1, 1, -1, -1\}} = |xyyxxxyyxx \dots\rangle
\end{align}

The sector-Neel state in this sector likewise exhibits fidelity oscillations, which are accurately captured by the forward-scattering approximation discussed in Sec.~\eqref{sec:FSA}. The corresponding dynamics of an illustrative example is shown in Fig.~\ref{fig:Mixedsectorfide}. We also consider some other examples in this work, which are given as a bond motif of $w=\{1, 1, -1, -1, 1, -1\}$.

\subsection{Forward Scattering Approximation}\label{sec:FSA}
We explain the fidelity oscillations using the forward scattering approximation (FSA)~\cite{Turner_2018_PXP}. We find that the structure of the FSA is universal across all three classes of sectors that exhibit QMBS. We first introduce the FSA Hamiltonian and then provide an intuitive picture of its action, explaining how it gives rise to coherent oscillations between the sector Neel state and the corresponding sector anti-Neel state. Throughout this discussion, we choose the sector Neel state as the state that begins with $|x\rangle$ as the reference state. This is just a convention, as the raising and lowering operators of the FSA Hamiltonian are defined relative to the chosen initial state.

Given the chosen initial state, we partition the sites of the chain into two subsets: $\mathcal{X}$, consisting of sites occupied by $|x\rangle$, and $\mathcal{Y}$, consisting of sites occupied by $|y\rangle$ in the initial sector-Neel state. In terms of this partition, the FSA-raising Hamiltonian is
\begin{equation}\label{eq:FSAHamiltonian}
\begin{aligned}
H^{+} &=
\sum_{i\in\mathcal{Y}}
\left(
|zz\rangle\langle yx|
-|zx\rangle\langle yz|
\right)_{i,i+1}  \\
&\quad+
\sum_{i\in\mathcal{X}}
\left(
|yx\rangle\langle zz|
-|yz\rangle\langle zx|
\right)_{i,i+1}.
\end{aligned}
\end{equation}

It is also apparent that $H^{-} = (H^{+})^{\dagger}$, and $H = H^{-} + H^{+}$, where $ H$ is the Hamiltonian Eq.~\eqref{eq:TranslationalinvariantKitaev}. Below we show how the above FSA Hamiltonian reduces to the familiar FSA Hamiltonians in the sectors with previously discovered scars in the $S=1$ Kitaev chain~\cite{You_scars_in_Kitaev_model, Balram_Kitaev_scars}.

\subsubsection{$0$ - wall sector}
In the PXP or $0$-wall sector, the initial Neel state is $|xyxy \dots\rangle$, and the corresponding FSA Hamiltonian is

\begin{equation}
\begin{aligned}
H^{+}={}&
\sum_{i\in\mathrm{odd}}
\left(
|yx\rangle\langle zz|
-|yz\rangle\langle zx|
\right)_{i,i+1} \\
&+
\sum_{i\in\mathrm{even}}
\left(
|zz\rangle\langle yx|
-|zx\rangle\langle yz|
\right)_{i,i+1}.
\end{aligned}
\end{equation}

It is important to note that configurations of the form $|yz\rangle$, $|zx\rangle$ never appear within the PXP sector. This follows from the constrained structure of the sector, i.e., $w_{j}=+1$ $\forall j$. This bond configuration implies that only $|zz\rangle$ and $|yx\rangle$ are possible on these bonds (see Appendix~\ref{app:Directorbasis}). Therefore, such configurations are forbidden, and the second term in the FSA Hamiltonian has no effect within this sector. 

This leaves the FSA Hamiltonian as, 
\begin{align}
    H^{+}= \sum_{i\in\mathrm{odd}} |yx\rangle\langle zz|_{i,i+1} + \sum_{i\in\mathrm{even}} |zz\rangle\langle yx|_{i,i+1}.
\end{align}

Using the mapping given in Ref.~\cite{Spin-1_Kitaev_model_Dhar} and Appendix of Ref.~\cite{Balram_Kitaev_scars}, the FSA Hamiltonian takes the form:
\begin{align}
   \nonumber H^{+} = \sum_{i \in \mathrm{odd}} |\downarrow \uparrow \downarrow\rangle \langle \downarrow\downarrow\downarrow| _{i-1/2, i+1/2, i+3/2} \\
    + \sum_{i \in \mathrm{even}} |\downarrow \downarrow \downarrow\rangle \langle \downarrow\uparrow\downarrow| _{i-1/2, i+1/2, i+3/2}
\end{align}

where $i - 1/2$ labels the bond between sites $i-1$ and $i$ (see the Appendix of Ref.~\cite{Balram_Kitaev_scars} for details), and the sector-Neel state maps to $|\downarrow\uparrow\downarrow\uparrow\cdots\rangle$. This precisely recovers the FSA Hamiltonian of the PXP model~\cite{Turner_2018_PXP}.

\subsubsection{$1$-bound wall sector}
In the $1$-bound wall sector, the sector-Neel state takes the form
$|xyx\rangle \equiv |xyxxyx\dots\rangle$, with
$\mathcal{X}=\{1,3,4,6,\dots\}$ and
$\mathcal{Y}=\{2,5,8,\dots\}$. The corresponding FSA Hamiltonian is
\begin{align}
    H^{+}
    &= \sum_{i \in \mathcal{Y}} |zz\rangle\langle yx|_{i,i+1}
    + \sum_{i \in \mathcal{X}}
    \left(
        |yx\rangle\langle zz|
        - |yz\rangle\langle zx|
    \right)_{i,i+1}.
\end{align}

The allowed processes are determined by the bond charges.
For $i\in\mathcal{Y}$, the bonds $(i, i+1)$ have $w_i=+1$, so only the first
term in Eq.~\eqref{eq:FSAHamiltonian} acts nontrivially.
The process $|zx\rangle\langle yz|_{i,i+1}$ vanishes on these bonds
because the configurations $|yz\rangle$ and $|zx\rangle$ are forbidden.
Within $\mathcal{X}$, the bonds indexed by $i\in\{1,4,7,\dots\}$
have $w_i=+1$ and support only the process
$|yx\rangle\langle zz|_{i,i+1}$, whereas those indexed by
$i\in\{3,6,9,\dots\}$ have $w_i=-1$ and support only the process
$-|yz\rangle\langle zx|_{i,i+1}$.
Thus, the FSA Hamiltonian reduces to
\begin{equation}\label{eq:FSA_balram_sector}
    \begin{aligned}
        H^{+}
        &= \sum_{i \in \{2,5,8,\dots\}}
        |zz\rangle\langle yx|_{i,i+1}
        + \sum_{i \in \{1,4,7,\dots\}}
        |yx\rangle\langle zz|_{i,i+1} \\
        &\quad - \sum_{i \in \{3,6,9,\dots\}}
        |yz\rangle\langle zx|_{i,i+1},
    \end{aligned}
\end{equation}
recovering the expression reported in
Ref.~\cite{Balram_Kitaev_scars}.

\begin{figure}
    \centering
    \begin{subfigure}{0.9\linewidth}
        \centering
        \includegraphics[width=\linewidth]{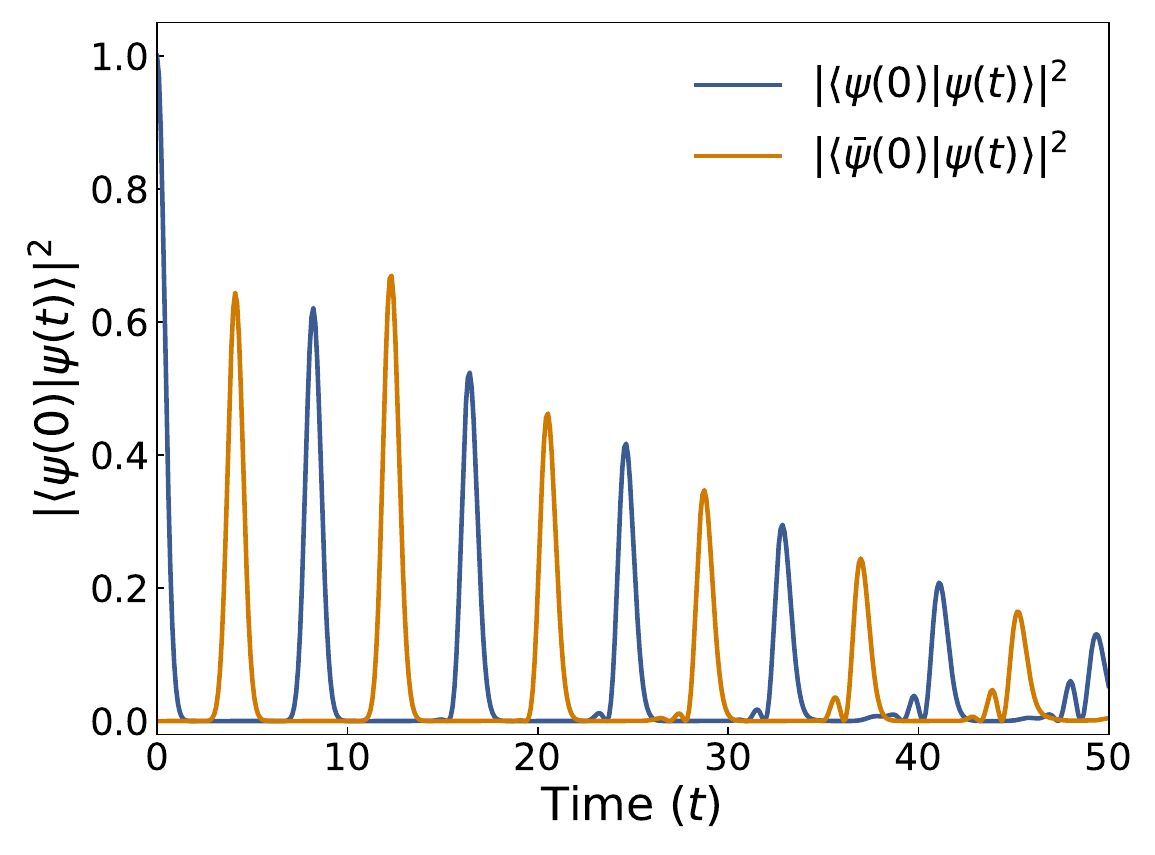}
        \caption{Fidelity.}
        \label{fig:3wallfide}
    \end{subfigure}

    \begin{subfigure}{\linewidth}
        \centering
        \includegraphics[width=\linewidth]{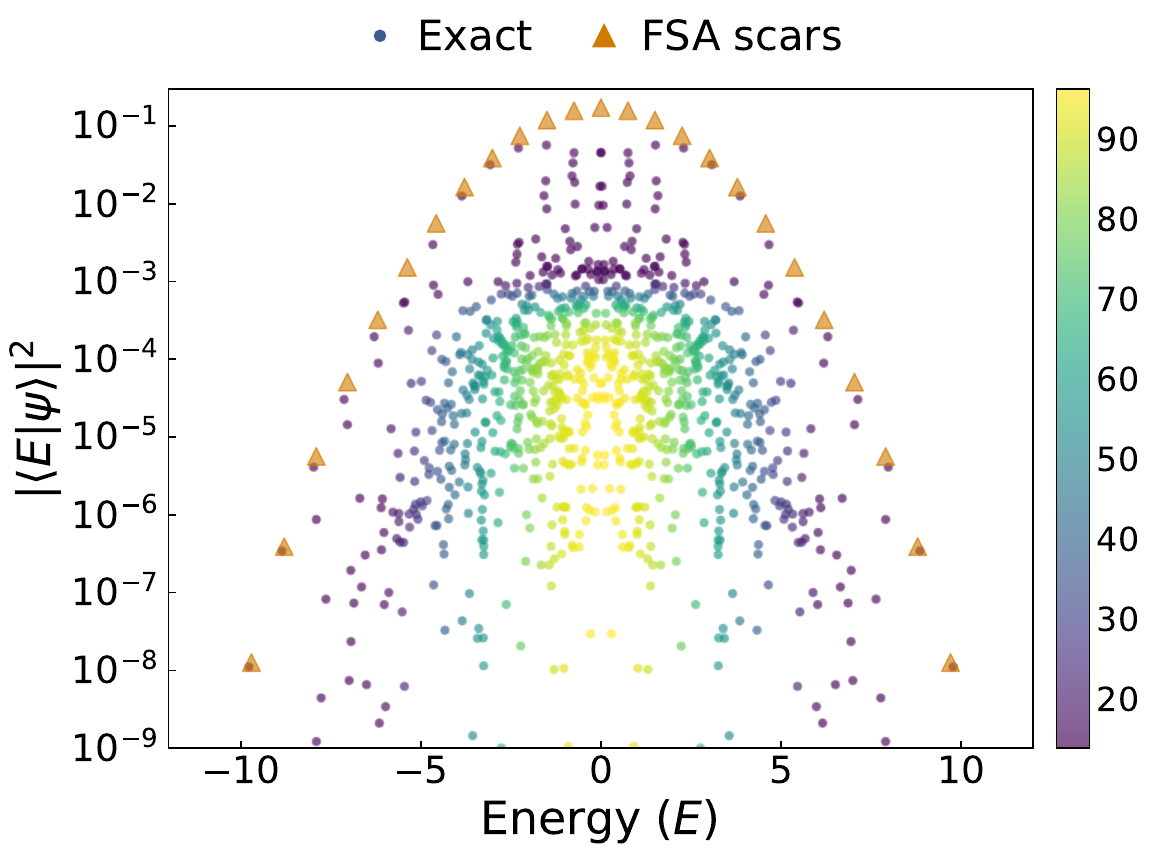}
        \caption{Overlap spectrum.}
        \label{fig:3walloverlap}
    \end{subfigure}
    \caption{Results for the $3$-wall sector for $L=24$ ($\dim=5778$). (a) Fidelity of the sector-Neel state $|\psi\rangle$ and the sector anti-Neel state $|\bar{\psi}\rangle$. (b) Overlap spectrum of the sector-Neel state.}
    \label{fig:3wall}
\end{figure}

\subsection{General $k$-wall periodic sector}
In this section, we present numerical evidence for PXP-like QMBS in periodic $k$-wall sectors that were previously unexplored. We focus on one
representative example, with results for several other values of $k$
provided in Appendix~\ref{app:kwallnumerics}.

We consider the sector with the repeating bond configuration
$w=\{-1,-1,-1,+1,-1,-1,-1,+1,\dots\}$. Following the construction
introduced in the previous section, the corresponding sector-Neel
state is
$|xxxxyyyy\rangle \equiv |xxxxyyyyxxxxyyyy\dots\rangle$.
We compute its exact dynamics under the Hamiltonian in
Eq.~\eqref{eq:TranslationalinvariantKitaev}. As shown in
Fig.~\ref{fig:3wall}, this state exhibits pronounced fidelity
revivals that are captured by the FSA.

\begin{figure}[!t]
    \centering
    \begin{subfigure}{0.9\linewidth}
        \centering
        \includegraphics[width=\linewidth]{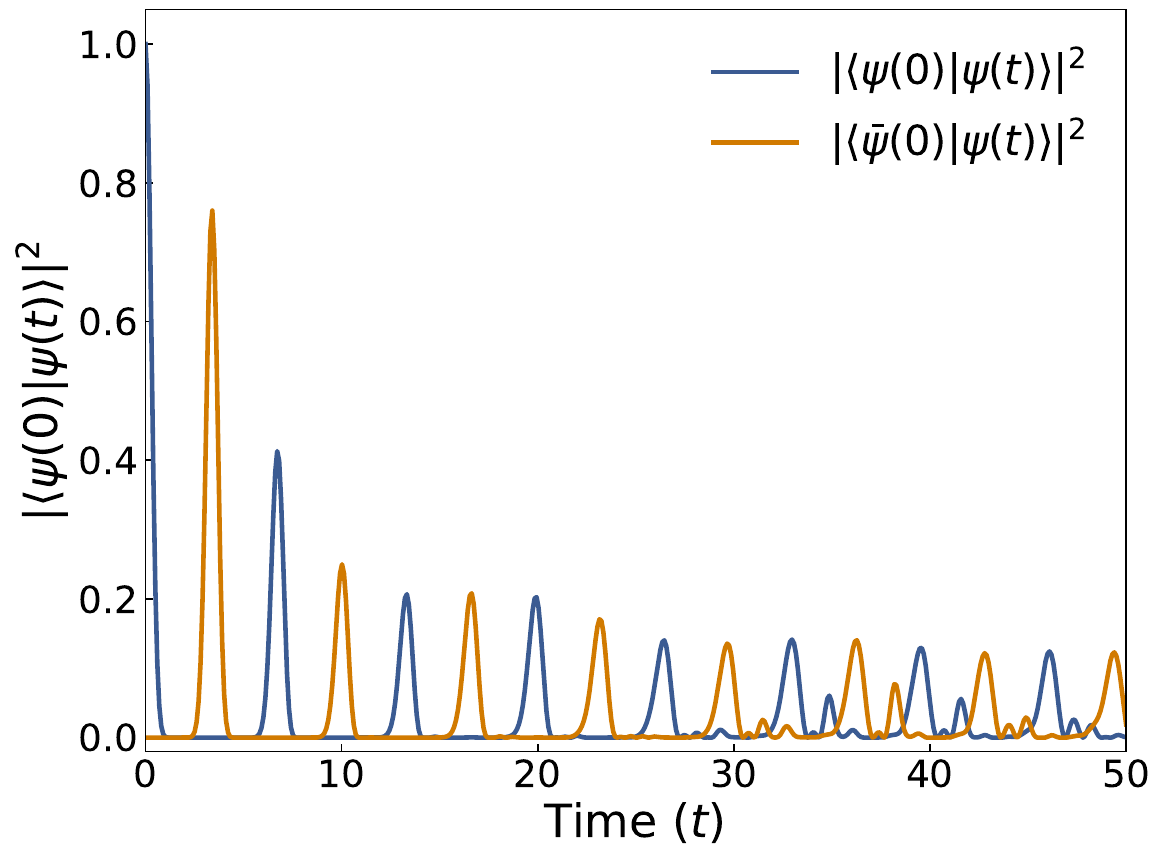}
        \caption{Fidelity.}
        \label{fig:2boundwallfide}
    \end{subfigure}

    \begin{subfigure}{0.9\linewidth}
        \centering
        \includegraphics[width=\linewidth]{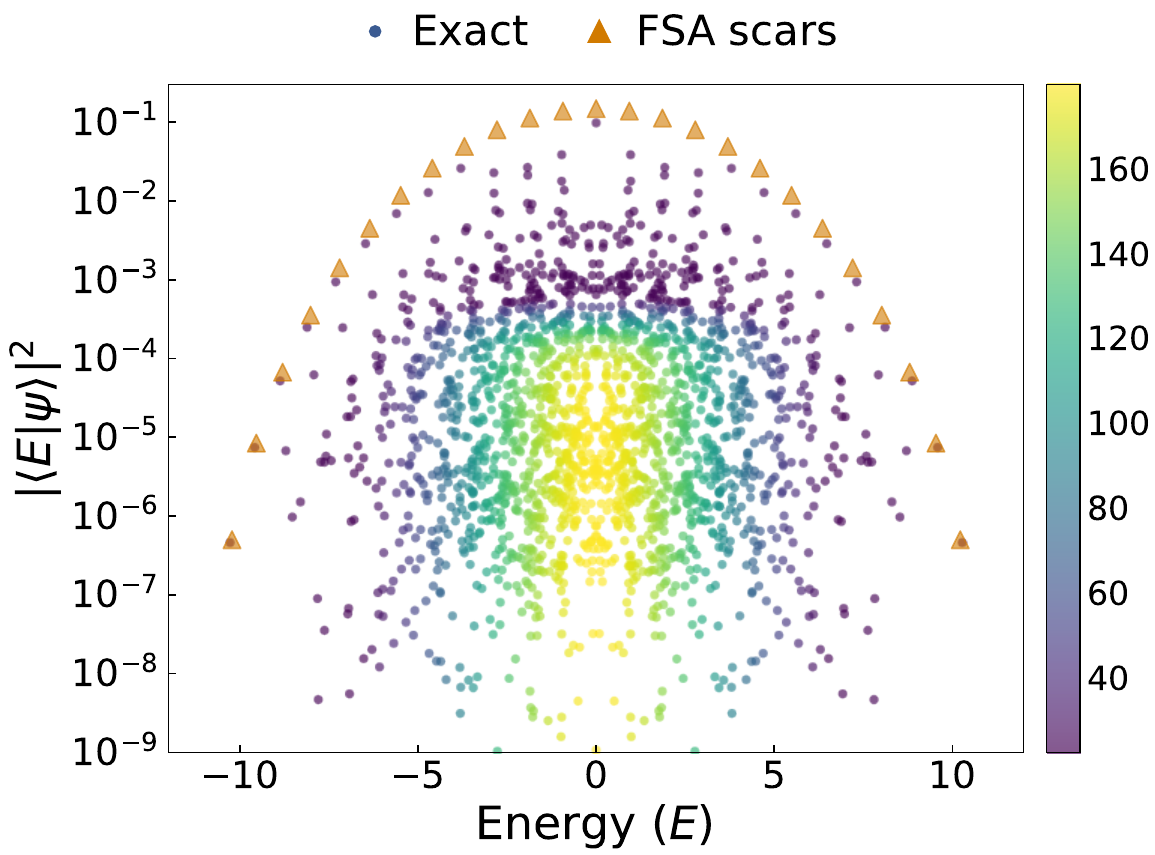}
        \caption{Overlap spectrum.}
        \label{fig:2boundwalloverlap}
    \end{subfigure}
    \caption{Results for the $2$-bound-wall sector for $L=24$ ($\dim=12098$). (a) Fidelity of the sector-Neel state $|\psi\rangle$ and the sector anti-Neel state $|\bar{\psi}\rangle$. (b) Overlap spectrum of the sector-Neel state.}
    \label{fig:2boundwall}
\end{figure}

\subsection{General $k$-bound-wall sectors}

We next turn to a second family of sectors, showing that the universal FSA description extends beyond the periodic $k$-wall sectors. As a representative example, we consider the $2$-bound-wall sector, with bond configuration
$w = \{+1,+1,-1,-1,\dots,+1,+1,-1,-1\}$ and repeating motif $(+1,+1,-1,-1)$; results for other values of $k$ are given in Appendix~\ref{app:boundwallnumerics}. Following the construction of Sec.~\ref{sec:sectorneel}, the corresponding sector-Neel state is
\begin{align}
|N\rangle_{\{1,1,-1,-1\}} = |xyxxxyxx\cdots\rangle .
\end{align}

As shown in Fig.~\ref{fig:2boundwallfide}, this state exhibits clear fidelity revivals, although they are weaker and decay faster than those in the $1$-bound-wall sector with motif $(+1,+1,-1)$. The overlap spectrum in Fig.~\ref{fig:2boundwalloverlap} shows the same trend: compared with the $1$-bound-wall sector, the FSA basis built from Eq.~\eqref{eq:FSAHamiltonian} captures the relevant exact eigenstates less accurately.

This sector was previously studied in Ref.~\cite{Balram_Kitaev_scars}, up to a lattice translation, where its weaker revivals were attributed to the reduced accuracy of the FSA. Our analysis is consistent with this interpretation and places it in a broader context. Within our framework, the $2$-bound-wall sector is described by the same universal FSA Hamiltonian, Eq.~\eqref{eq:FSAHamiltonian}, that accurately captures the strongly scarred sectors. Its reduced accuracy here therefore cannot be attributed to a sector-specific choice of FSA construction. Instead, it points to an intrinsic property of the sector, which we interpret as a stronger breaking of the approximate SU(2) structure underlying the scarred dynamics~\cite{Weakly_broken_lie_algebra_pxp}.

\subsection{Mixed periodic sectors}
In this section, we show that the construction of the sector-Neel state and the FSA Hamiltonian introduced above also apply to the mixed sectors defined in Sec.~\ref{sec:Mixedsector}. As a representative example, we consider the sector with bond configuration $w=\{+1,+1,-1,+1,-1,\dots\}$, whose corresponding sector-Neel state is given in Eq.~\eqref{eq:mixedsectorneel}. Results for additional mixed sectors are presented in Appendix~\ref{app:mixedsectornumerics}.

\begin{figure}
    \centering
    \includegraphics[width=0.9\linewidth]{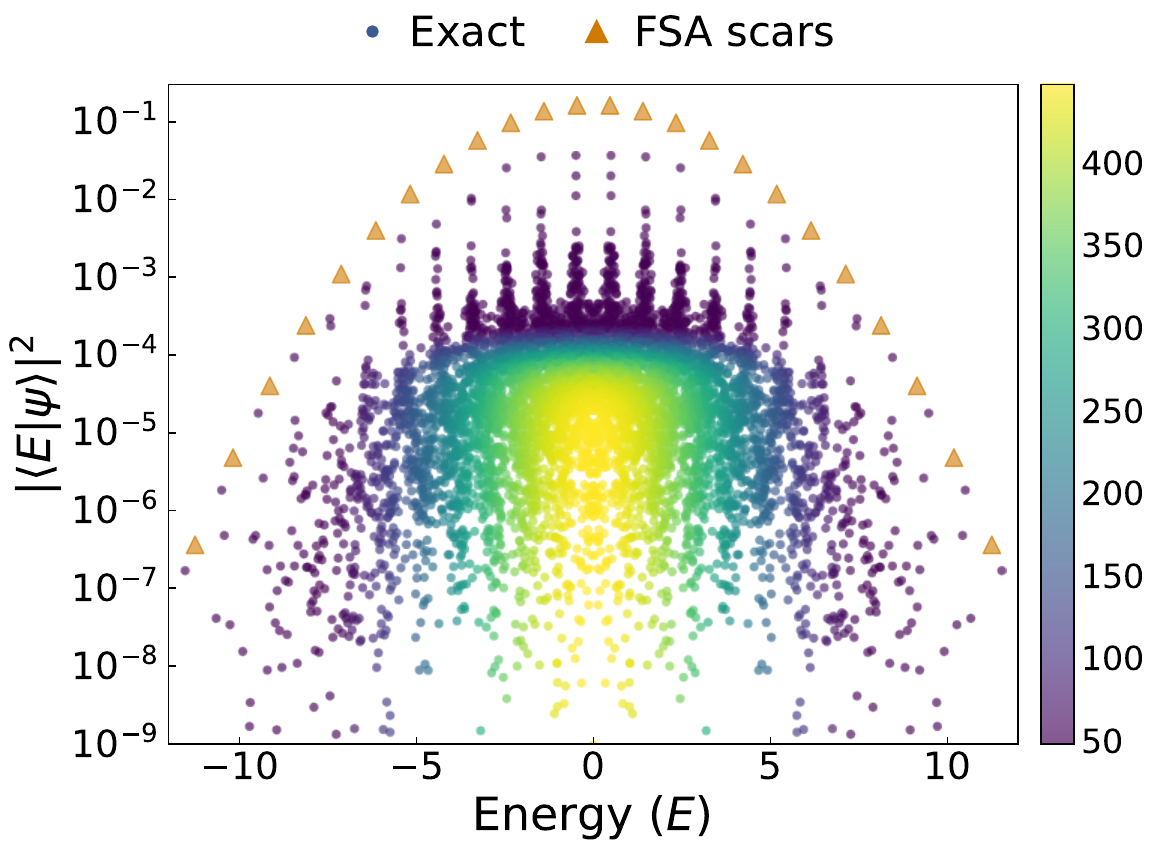}
    \caption{Overlap spectrum for the sector-Neel state for mixed sector $w=\{1, -1, 1, -1, -1, \dots \}$ for $L=25$}
    \label{fig:Mixedsectoroverlap}
\end{figure}

As shown in Fig.~\ref{fig:Mixedsectorfide}, the sector-Neel state exhibits fidelity revivals. The corresponding overlap spectrum, shown in Fig.~\ref{fig:Mixedsectoroverlap}, further supports the ability of the FSA Hamiltonian to capture the dynamics in this sector.

\begin{figure*}
    \includegraphics[width=0.7\linewidth]{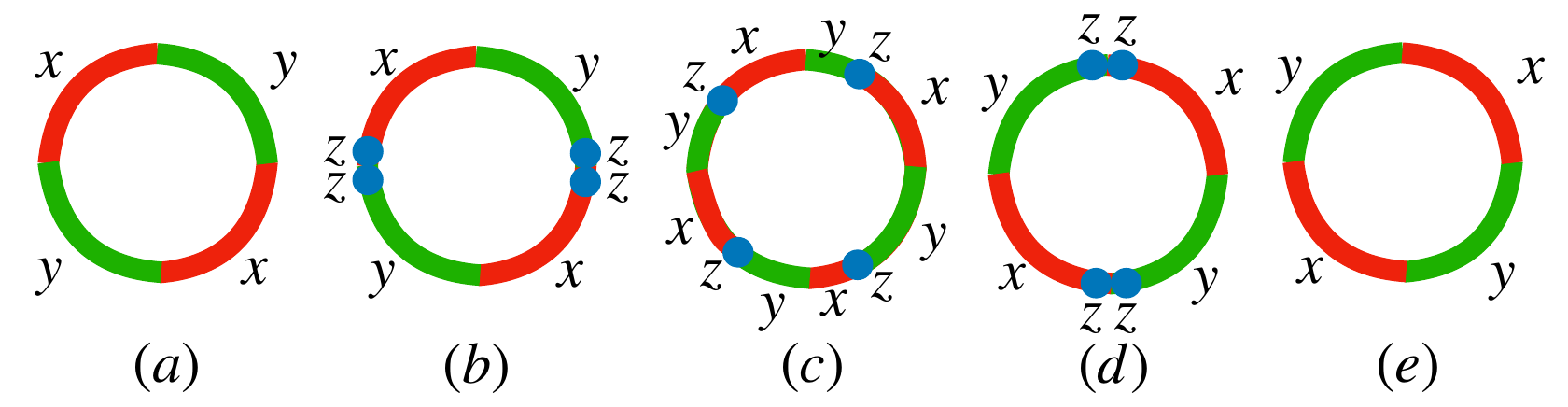}
    \caption{Schematic illustration of the slider analogy; it shows the possible processes that can happen, not a unique trajectory of the wavefunction. Red and green arcs represent the $|x\rangle$ and $|y\rangle$ Cartesian strings, respectively, while blue points represent defects corresponding to $|z\rangle$ states. Walls lie at the interfaces between red and green arcs; the example shown contains four walls. Panels (a) and (e) depict the two sector-Neel states. Starting from (a), defects are created at the appropriate walls in (b), propagate in (c), and annihilate in pairs in (d), yielding the other sector-Neel state in (e).}
    \label{fig:DefectSlider}
\end{figure*}

\subsection{A physical picture}\label{subsec:slider}
The dynamics generated by the FSA Hamiltonian $H^{+}$ in
Eq.~\eqref{eq:FSAHamiltonian} within a given bond-configuration
sector can be understood through a physical picture illustrated
in Fig.~\ref{fig:DefectSlider}. Consider an initial defect-free
configuration in a four-wall sector, consisting of alternating
arcs of $|x\rangle$ and $|y\rangle$ states. The walls lie at
the interfaces between these arcs. Pairs of $|z\rangle$
states can be created at the two walls where a $|y\rangle$
arc lies to the left of an $|x\rangle$ arc, according to the
rules in Eq.~\eqref{app:eq:+1_rules}.

These $|z\rangle$ states act as mobile defects: each defect
propagates freely within its respective arc, converting
$|x\rangle$ states into $|y\rangle$ states and vice versa.
This motion can be visualized as a slider, with black dots
representing the $|z\rangle$ defects moving through the arcs
and exchanging green $|y\rangle$ states with red
$|x\rangle$ states. Throughout this motion, the ordering
constraint is maintained: green $|y\rangle$ states lie to
the left of the black $|z\rangle$ defects, while red
$|x\rangle$ states lie to their right. Once the defects
have traversed their arcs, they can annihilate pairwise upon
arriving simultaneously at a wall.

These dynamical rules cause the system to oscillate between
two defect-free configurations consisting of alternating
$|x\rangle$ and $|y\rangle$ arcs, related by interchanging
$|x\rangle$ and $|y\rangle$. This picture applies to every
even-wall sector. In contrast, odd-wall sectors admit no
defect-free configuration: at least one $|z\rangle$ defect
must always be present in one of the arcs.

\section{Sectors without pronounced revivals from defect-free states}
\label{sec:NonScarredSectors}
\begin{figure}[H]
    \centering
    \includegraphics[width=0.9\linewidth]{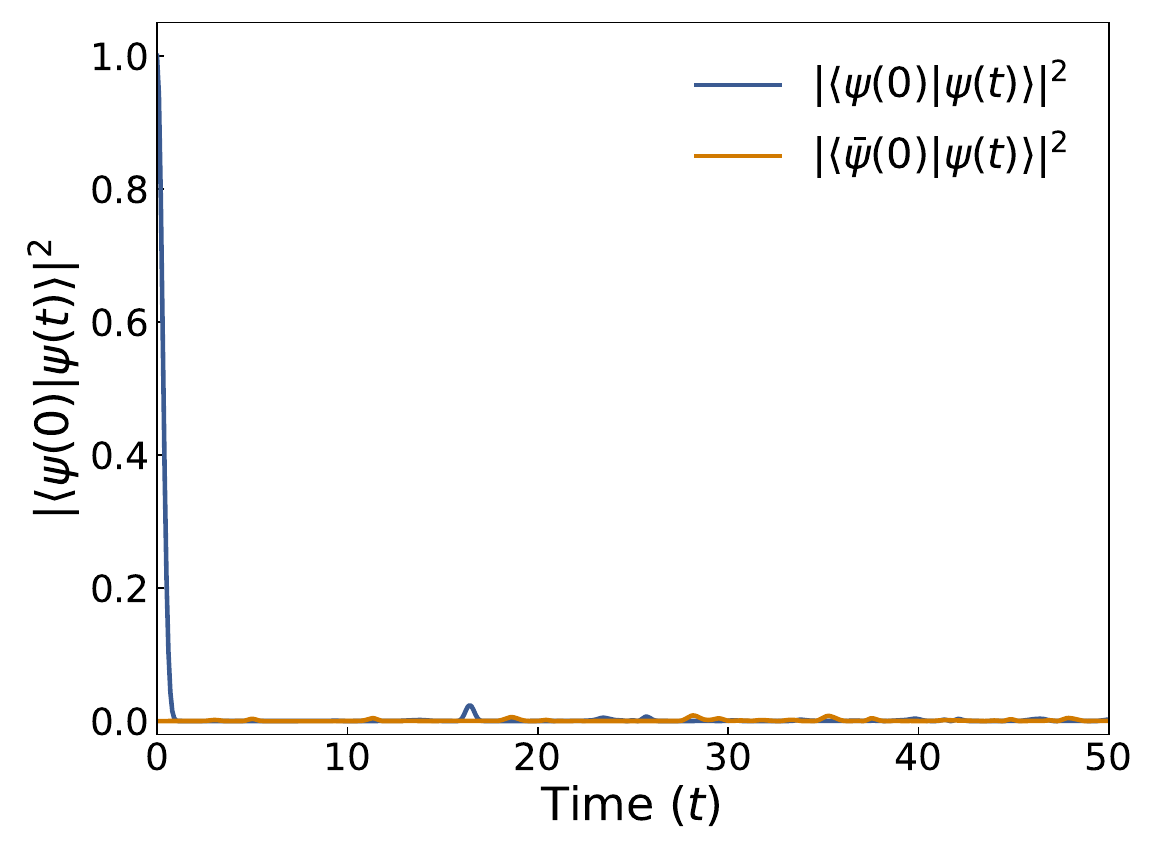}
    \caption{Fidelity dynamics, for sector $w=\{1, 1, 1, -1, -1, -1\dots\}$, for L=24 with 10402 states in the sector, $|\psi\rangle$ represents the sector-Neel state, while $|\bar{\psi}\rangle$ representing the sector-Anti-Neel state}
    \label{fig:111-1-1-1sector}
\end{figure}

The construction of the sector-Neel state presented in Sec.~\ref{sec:SectorResolvedApproximateScarStudy} applies to any sector with an even number of walls. However, the existence of this state and the associated FSA Hamiltonian does not by itself imply scarred dynamics. To illustrate this, we consider the sector $w=\{1,1,1,-1,-1,-1,\dots,1,1,1,-1,-1,-1\}$, for which our construction yields the sector-Neel state $|xyxyyyyxyxxxx\dots\rangle$. This state exhibits no fidelity revivals, as shown in Fig.~\ref{fig:111-1-1-1sector}. Similar behavior is observed in other sectors containing three or more consecutive walls, with supporting numerical evidence provided in Appendix~\ref{app:Nonscarred_Sectors}.

\begin{figure*}
    \centering
    \begin{subfigure}{0.49\linewidth}
        \centering
        \includegraphics[width=\linewidth]{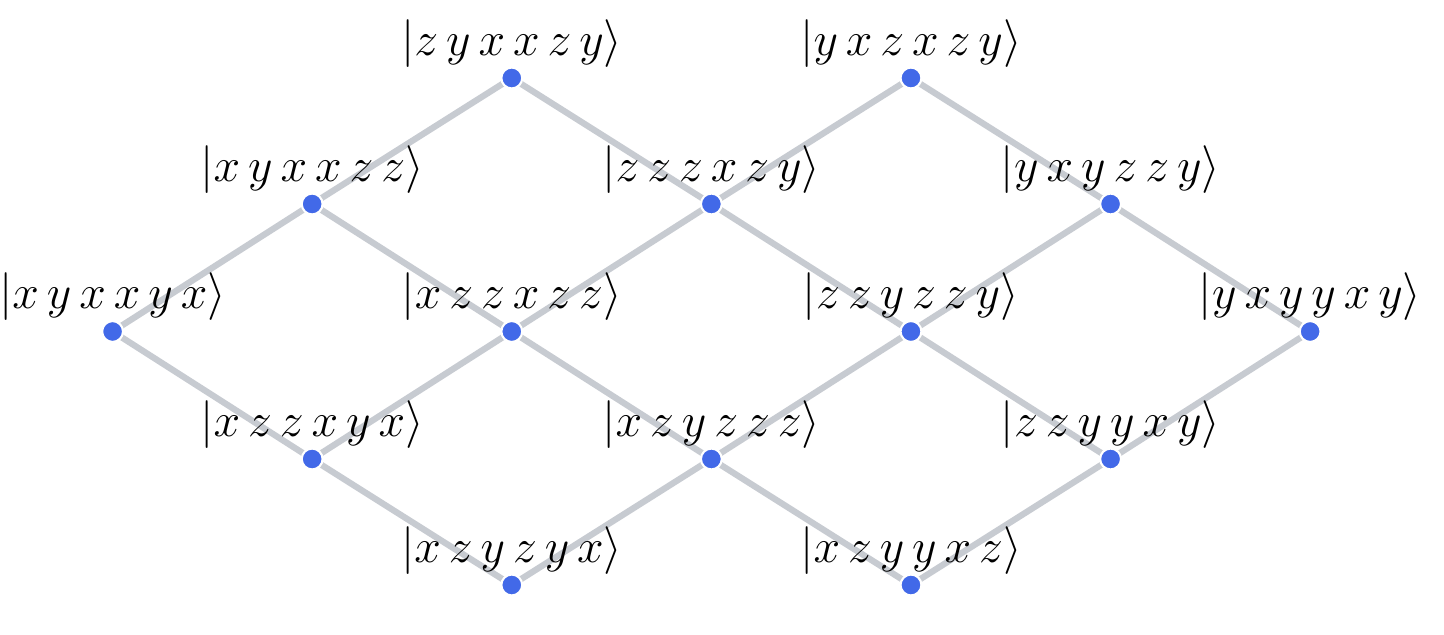}
        \caption{Sector w=\{1, 1, -1, 1, 1, -1\}}
        \label{fig:graphbalram}
    \end{subfigure}\hfill
    \begin{subfigure}{0.49\linewidth}
        \centering
        \includegraphics[width=\linewidth]{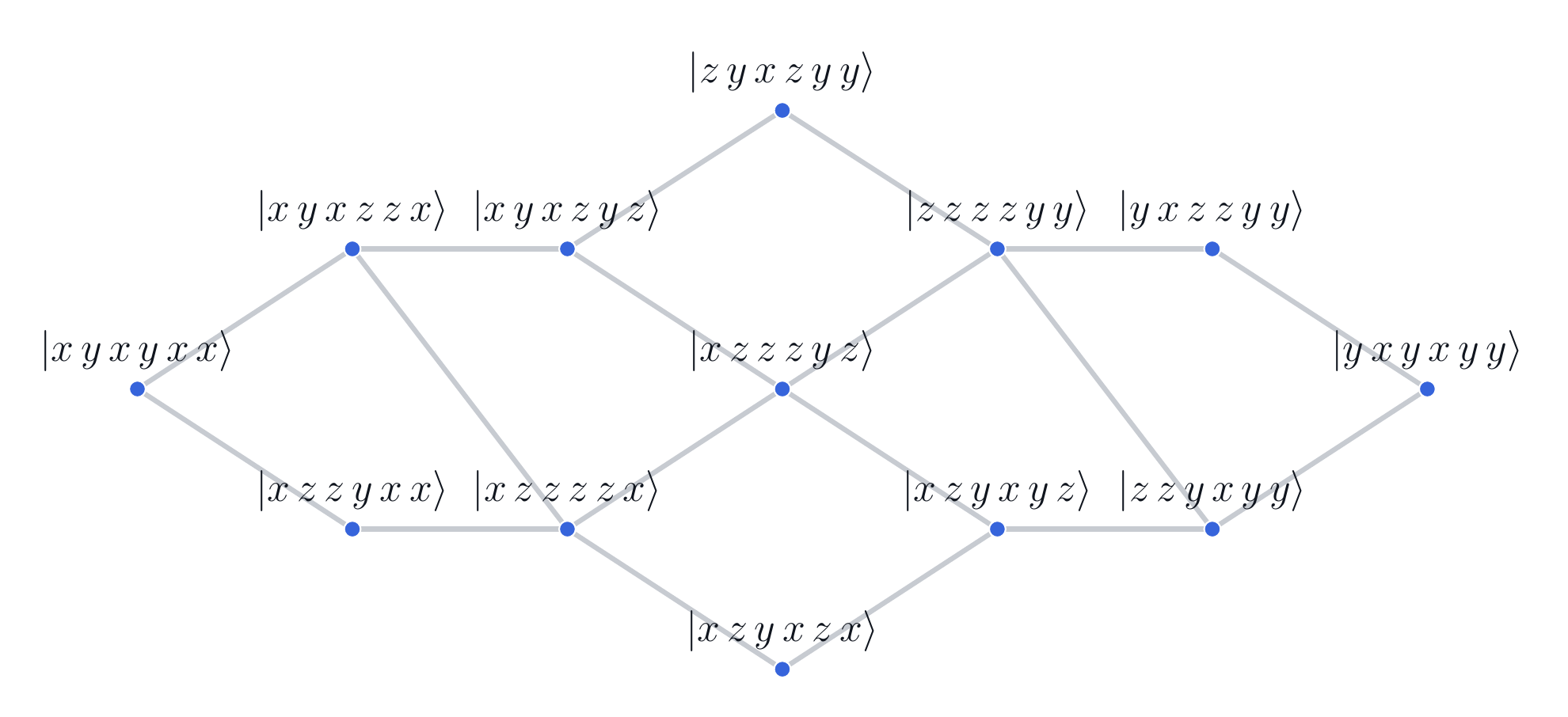}
        \caption{Sector w=\{1, 1, 1, 1, -1, -1\}}
        \label{fig:graphw1111-1-1}
    \end{subfigure}
    \caption{Connectivity graphs of the Hilbert space for two different sectors for L=6 system size. The Hilbert space graph is symmetric with respect to Neel and anti-Neel nodes for graph (a), while this symmetry doesn't exist in this Hilbert space graph, which inhibits state transfer.}
    \label{fig:combined_graphs}
\end{figure*}
The physical origin of this behavior can be understood heuristically in terms of defect dynamics. Any bond configuration decomposes into \emph{wall clusters}, i.e., maximal runs of consecutive walls, separated by \emph{arcs} of no-wall bonds. Defect pairs are created and annihilated on walls [Eq.~\eqref{app:eq:-1_rules}, \eqref{app:eq:+1_rules}], but can propagate only along arcs, since a defect cannot cross a wall. Two elementary processes therefore arise. A pair whose defects border an arc can \emph{slide}: its defects traverse the neighboring arcs and annihilate at the opposite walls, as in the slider picture of Sec.~\ref{subsec:slider}. A pair created on an interior bond of a cluster of three or more consecutive walls cannot propagate and can only \emph{blink}, being annihilated and re-created in place.

Neither process alone precludes scarring. In the PXP sector, $w_j = +1$ for all $j$, all pairs blink, and the dynamics is the well-known PXP scarring. In the $k$-wall, $k$-bound-wall, and mixed sectors, every wall cluster contains at most two walls, so every pair slides. In both cases, the elementary processes are homogeneous, and the local transfers between the sector-Neel and sector-anti-Neel configurations remain in phase. In sectors containing a cluster of three or more walls together with arcs, blinking and sliding pairs coexist and evolve on different timescales, which dephases the global transfer. For example, in the sector $w = \{1,1,1,-1,1,1,1,-1\}$ at $L = 8$,
\begin{align*}
H^+|xyxyyxyx\rangle = |x\underbrace{zz}_{\text{blinking}}yyxyx\rangle \\
+ |xyxyzzyx\rangle + |xyxyyxzz\rangle ,
\end{align*}
where the first pair can only blink while the other two slide. All sectors in Fig.~\ref{app:fig:nonscar_diagnostics} share this structure. The same reasoning suggests that, within the sliding class, unequal arc lengths in a motif, which imply unequal transit times, weaken revivals.

We stress that this is a heuristic for when the slider mechanism produces coherent transfer from the sector-Neel state, not a strict criterion: some sectors with mixed processes show isolated revival peaks (Fig.~\eqref{app:fig:nonscar_diagnostics}), and revival quality within the sliding class also depends on arc length (Appendix~\ref{app:boundwallnumerics}).

This obstruction is also reflected in the Hilbert-space connectivity graph, which lacks symmetry under exchange of the sector-Neel and sector-anti-Neel nodes. To illustrate this, we compare the sectors $w=\{1,1,1,1,-1,-1\}$ and $w=\{1,1,-1,1,1,-1\}$ at $L=6$, as shown in Fig.~\ref{fig:combined_graphs}. For the sector containing more than two consecutive walls, the graph no longer exhibits this symmetry. We argue that its absence obstructs coherent transfer between the two extremal states, thereby suppressing scarred dynamics arising from the slider mechanism.

The numerical results in Appendix~\ref{app:Nonscarred_Sectors} suggest that sectors containing more than two consecutive walls generally lack scarred dynamics from the sector-Neel state, consistent with the slider analogy. Fig.~\ref{app:fig:mixed4_wall_diagnostics} illustrates this behavior in a mixed-wall sector. However, the absence of fidelity revivals from this initial state does not rule out QMBS in the sector: as the FSA construction depends on the choice of initial state, and other product states or states with low entanglement may exhibit scarred dynamics through different mechanisms. Conversely, revivals observed in some sectors with more than two consecutive walls, if they exist, may reflect mechanisms beyond the slider analogy. A systematic understanding of these alternative mechanisms and the sectors that support them remains an open question.

\section{Discussion}
\label{sec:Discussion}

We studied features of the spectrum of the $S=1$ Kitaev chain with the help of its Hilbert-space connectivity structure. Using a transfer-matrix formalism, we derived the chiral index, defined as the number of zero modes enforced by sublattice imbalance in each bond-charge sector, and showed that its sum across different sectors for even-length $L$ periodic chains is given by the Jacobsthal numbers $J_{L+1}$. In the presence of weak bond disorder that removes the spatial symmetries, this bound is saturated in every sector for all system sizes examined. Building on recent constructions in the PXP model~\cite{Volume_law_scar_PXP} and other spin chains~\cite{Exact_volume_law_zero_modes}, we obtained exact analytic zero-energy eigenstates with antipodal correlations in sectors invariant under translation by $L/2$. We showed the existence of $2^{L/2}$ such states.
For contiguous bipartitions these states have Von Neumann entanglement entropy $S = \ln D_w$, where $D_w$ is the sector dimension for a bond configuration $\vec{w}$. In sectors whose dimension grows exponentially with system size, this entropy follows the conventional volume law. 

Using the Choi operator--state mapping, we gave an alternative proof of their existence that makes explicit the conditions on the chiral symmetry and on the Hamiltonian under which the construction applies. Since the proof uses the on-site structure of $\mathcal{C}$ only through the transfer relation in Eq.~\eqref{eq:push_through}, the construction may extend beyond on-site, invertible chiral symmetries. For symmetries represented by matrix product operators (MPO), including non-invertible ones, the transfer relation would be replaced by an MPO pull-through identity. Where such an identity exists, it would define an effective transfer map that plays the same role in the seam construction. We leave this extension to future work.

We further identified three classes of bond-charge sectors, $k$-wall periodic, $k$-bound-wall, and mixed periodic, in which the sector-Neel state exhibits fidelity revivals. Within the representative sectors studied, these revivals are approximately captured by an FSA Hamiltonian of a common form, which reduces to the known FSA descriptions in the PXP sector~\cite{You_scars_in_Kitaev_model} and in the other sectors studied in Ref.~\cite{Balram_Kitaev_scars}. The quality of both the revivals and the FSA description varies considerably across sectors. To get some understanding of the fidelity revivals, we proposed a heuristic physical picture, in which defect pairs are created at walls, propagate along arcs, and annihilate pairwise. In this picture, revivals are expected when the elementary defect processes are homogeneous, as in the PXP sector, where all pairs blink in place, and in the scarred classes, where all pairs propagate. Sectors in which blinking and propagating pairs coexist typically show no revivals from the sector-Neel state. 

Although the physical picture discussed above accounts for the behavior of many sectors, it does not provide a complete criterion for scarred dynamics. Some sectors exhibit revivals despite the obstructions suggested by this picture, whereas others fail to exhibit revivals even when coherent defect propagation appears possible. These exceptions suggest that additional features of the Hilbert-space connectivity and the associated transition amplitudes may influence the dynamics. Moreover, the absence of revivals from a sector-Neel state does not exclude QMBS associated with other initial states. Understanding these exceptions and identifying alternative scarring mechanisms remains an open question. A similar picture has been reported in a different setting. In Ref.~\cite{Sanjay_scar_landau_levels}, the effective Hamiltonian of a Landau level on a thin torus maps exactly onto the PXP model at filling fraction $\nu = 1/3$. Yet, PXP-like scarring persists at other filling fractions. Our results are analogous: scarring is not confined to the sector that maps onto the PXP model but extends to a broad family of other sectors.

Another important direction for future work is to characterize the Hilbert-space connectivity graphs of these sectors and to identify the structures that enable coherent state transfer. Since the defect dynamics can be viewed as tight-binding motion supplemented by pair creation and annihilation at walls, it is natural to look for graph symmetries and connectivity patterns that underlie the scarred dynamics, following the approach of Ref.~\cite{Zlatko_Hypergrid_origin}. Such an analysis would also clarify the limits of the physical picture. The adjacency-matrix representation of the Kitaev chain Hamiltonian obtained through an appropriate unitary transformation in Ref.~\cite{pandey2026exactrokhsarkivelsonpointshosting} may provide a direction for relating this connectivity to spectral properties. In particular, the zero modes observed in our simulations arose from an imbalance between the two sublattices of the Hilbert-space graphs associated with the connected components. However, irregular bipartite graphs can, in general, support additional zero modes arising from local imbalances, even when the two sublattices are globally balanced~\cite{Saynal_etal_PRL_2016,Bhola_etal_PRX_2022,Nicolau_etal_PRX}. However, we did not observe any such zero modes in our simulations. Understanding the origin of their absence and determining whether this feature persists more generally would be another interesting direction for future investigation.

The close resemblance between the scars identified here and those in the PXP model also raises the question of whether they admit a description in terms of quasiparticle condensation, as proposed in Ref.~\cite{QMBS_frommagnoncondensation}, with the $|zz\rangle$ defect pairs serving as candidate quasiparticles, which could also explain the presence of secondary towers seen in many sectors. Another promising direction is to investigate whether lattice decorations, similar to those introduced in Ref.~\cite{verde2026engineeringquantummanybodyscars}, can mobilize otherwise immobile defects, restore coherent state transfer, and stabilize scarred dynamics in sectors that lack fidelity revivals. Finally, extending this study to higher-dimensional Kitaev and related compass models~\cite{Compass_model_review,pandey2026exactrokhsarkivelsonpointshosting} may reveal how lattice geometry affects defect mobility, zero-mode structure, and quantum many-body scarring.

\begin{acknowledgments}
The authors thank Vikram Tripathi, Kedar Damle, and Sanjay Moudgalya for helpful comments on an earlier version of this manuscript and Vikram Tripathi and Kedar Damle for their guidance and encouragement. Claude Opus 5.5 was used to assist in writing the plotting scripts and polish the language in parts of the manuscript. This work was supported by the Department of Atomic Energy, Government of India, under Project Identification No. RTI 4002. We also thank Kapil Ghadiali and Ajay Salve for technical support. Numerical simulations of the spin systems were performed using the Python library QuSpin~\cite{Quspin_2017}, together with custom codes based on Ref.~\cite{Anders_computational_spin_systems} and matrix-free methods described in Ref.~\cite{kahanamoku-meyer_exploring_2023}. The authors take full responsibility for all results and claims presented in this work.
\end{acknowledgments}

\FloatBarrier
\bibliography{bibliography}

\appendix
\section{Spin-1 Algebra basics}\label{app:Directorbasis}

Throughout this work, $|-1\rangle$, $|0\rangle$, and $|1\rangle$ denote the eigenstates of $S^{z}$ with eigenvalues $-1$, $0$, and $1$, respectively. We define the Cartesian basis as
\begin{align}
    |x\rangle &= \frac{|-1\rangle-|1\rangle}{\sqrt{2}}, \\
    |y\rangle &= \frac{i\bigl(|-1\rangle+|1\rangle\bigr)}{\sqrt{2}}, \\
    |z\rangle &= |0\rangle.
\end{align}
In this basis, the action of the spin-1 operators takes the compact form
\begin{equation}
    S^{\alpha}|\beta\rangle
    = i\sum_{\gamma}\epsilon_{\alpha\beta\gamma}|\gamma\rangle,
\end{equation}
where $\epsilon_{\alpha\beta\gamma}$ is the Levi-Civita symbol. In the ordered basis $\{|x\rangle,|y\rangle,|z\rangle\}$, the corresponding matrix representations are
\begin{align}\label{app:eq:spin_op_matrix}
S^x &=
\begin{pmatrix}
0 & 0 & 0 \\
0 & 0 & -i \\
0 & i & 0
\end{pmatrix},
&
S^y &=
\begin{pmatrix}
0 & 0 & i \\
0 & 0 & 0 \\
-i & 0 & 0
\end{pmatrix},
&
S^z &=
\begin{pmatrix}
0 & -i & 0 \\
i & 0 & 0 \\
0 & 0 & 0
\end{pmatrix}.
\end{align}
The parity operators therefore take the diagonal forms
\begin{equation}
\begin{aligned}
\mathcal{P}^x &= \operatorname{diag}(1,-1,-1),\\
\mathcal{P}^y &= \operatorname{diag}(-1,1,-1),\\
\mathcal{P}^z &= \operatorname{diag}(-1,-1,1).
\end{aligned}
\end{equation}

These expressions imply
\begin{equation}
    [\mathcal{P}^{\alpha},S^{\alpha}]=0,
    \qquad
    \{\mathcal{P}^{\alpha},S^{\beta}\}=0
    \quad (\alpha\neq\beta).
\end{equation}

We next identify the local two-site basis states compatible with each eigenvalue of the conserved bond operator. For $w_j=-1$, the allowed states are
\begin{equation}
    |xx\rangle,\quad |yy\rangle,\quad |yz\rangle,\quad |zx\rangle,
\end{equation}
whereas for $w_j=+1$, they are
\begin{equation}
    |xy\rangle,\quad |xz\rangle,\quad |zy\rangle,\quad
    |zz\rangle,\quad |yx\rangle.
\end{equation}
Using Eq.~\eqref{app:eq:spin_op_matrix}, the action of the local Hamiltonian term $S^x_jS^y_{j+1}$ in the $w_j=-1$ sector is
\begin{equation}\label{app:eq:-1_rules}
\begin{aligned}
    S^x_jS^y_{j+1}|xx\rangle &= 0, \\
    S^x_jS^y_{j+1}|yy\rangle &= 0, \\
    S^x_jS^y_{j+1}|yz\rangle &= -|zx\rangle, \\
    S^x_jS^y_{j+1}|zx\rangle &= -|yz\rangle.
\end{aligned}
\end{equation}
Similarly, in the $w_j=+1$ sector,
\begin{equation}\label{app:eq:+1_rules}
\begin{aligned}
    S^x_jS^y_{j+1}|xy\rangle &= 0, \\
    S^x_jS^y_{j+1}|xz\rangle &= 0, \\
    S^x_jS^y_{j+1}|zy\rangle &= 0, \\
    S^x_jS^y_{j+1}|zz\rangle &= |yx\rangle, \\
    S^x_jS^y_{j+1}|yx\rangle &= |zz\rangle.
\end{aligned}
\end{equation}

Each nonzero transition connects basis states with opposite chiral index, since the parity of the count of $|z\rangle$ states on odd sites changes under the action of the Hamiltonian. As a consequence, the Hamiltonian maps the chiral-index $+1$ subspace into the chiral-index $-1$ subspace, and vice versa. The Hilbert-space connectivity graph is consequently bipartite, with the two sublattices distinguished by their chiral charges. An imbalance between the dimensions of these subspaces enforces zero modes, as discussed in the main text.

\section{Full Zero mode counting}\label{app:full_zm_count}
In this appendix, we derive the total number of zero modes enforced by chiral sublattice imbalance. For an even-length chain, it is convenient to group the transfer matrices into pairs and define
\begin{align}
    B_1 &= T^{(-)}DT^{(-)}, &
    B_2 &= T^{(-)}DT^{(+)}, \\
    B_3 &= T^{(+)}DT^{(-)}, &
    B_4 &= T^{(+)}DT^{(+)}.
\end{align}
The imbalance in Eq.~\eqref{eq:imbalance_eqn} can then be expressed as the trace of a product of these paired matrices.

We regard $B_1,\ldots,B_4$ as letters of an alphabet, so that each bond-charge configuration corresponds to a word. Although the number of words grows exponentially with system size, their matrix products belong to a finite set: every nonempty product reduces, up to an overall sign, to one of six representatives $M_1,\ldots,M_6$. These are
\begin{equation}\label{eq:newtransfer_matrices}
\begin{aligned}
M_1 &=
\begin{pmatrix}
 1 & 0 & 0\\
-1 & 1 & 1\\
 1 & 0 & 0
\end{pmatrix},
&\qquad
M_2 &=
\begin{pmatrix}
0 &  1 &  1\\
1 & -1 & -1\\
0 &  1 &  1
\end{pmatrix},
\\[6pt]
M_3 &=
\begin{pmatrix}
-1 & 1 & 1\\
 1 & 0 & 0\\
-1 & 1 & 1
\end{pmatrix},
&\qquad
M_4 &=
\begin{pmatrix}
1 & -1 & -1\\
0 &  1 &  1\\
1 & -1 & -1
\end{pmatrix},
\\[6pt]
M_5 &=
\begin{pmatrix}
0 & -1 & -1\\
1 &  0 &  0\\
0 & -1 & -1
\end{pmatrix},
&\qquad
M_6 &=
\begin{pmatrix}
1 &  0 &  0\\
0 & -1 & -1\\
1 &  0 &  0
\end{pmatrix}.
\end{aligned}
\end{equation}
with
\begin{equation}
    M_\alpha=B_\alpha,
    \qquad \alpha=1,\ldots,4.
\end{equation}
Their multiplication table is
\begin{equation}\label{eq:Truth_table}
\begin{array}{c|cccccc}
M_iM_j & M_1 & M_2 & M_3 & M_4 & M_5 & M_6 \\ \hline
M_1 & M_1 & M_2 & M_3 & M_4 & M_5 & M_6 \\
M_2 & M_2 & M_1 & -M_5 & M_6 & -M_3 & M_4 \\
M_3 & M_3 & M_4 & M_1 & M_2 & M_6 & M_5 \\
M_4 & M_4 & M_3 & -M_6 & M_5 & -M_1 & M_2 \\
M_5 & M_5 & -M_6 & -M_2 & -M_1 & -M_4 & M_3 \\
M_6 & M_6 & -M_5 & -M_4 & -M_3 & -M_2 & M_1
\end{array}.
\end{equation}
Thus, the set is closed under multiplication up to an overall sign. After identifying matrices that differ only by a sign, the multiplication structure is isomorphic to the permutation group $S_3$, with $M_1$ acting as the identity within this set.

This finite closure is the key to the counting problem. The imbalance $\Delta_{\vec{w}}$ is signed, whereas the number of zero modes it enforces is $|\Delta_{\vec{w}}|$. Summing the signed imbalances over sectors would therefore introduce cancellations. Instead, we count the configurations associated with each matrix class and weight them by the absolute value of its trace. Overall signs can be discarded because they do not affect this weight.

For a chain of length $2L$, let $a^\alpha_{2L}$ denote the number of bond-charge configurations whose matrix product equals $\pm M_\alpha$. The total number of imbalance-enforced zero modes is
\begin{equation}
    N_{2L}
    = \sum_{\alpha=1}^{6}
    a^\alpha_{2L}\left|\operatorname{Tr}M_\alpha\right|.
\end{equation}
We collect these configuration counts into the row vector
\begin{equation}
    \mathbf{x}_{2L}
    =
    \begin{pmatrix}
    a^1_{2L} & a^2_{2L} & a^3_{2L} &
    a^4_{2L} & a^5_{2L} & a^6_{2L}
    \end{pmatrix}.
\end{equation}
Appending a pair of bond charges corresponds to right multiplication by one of $B_1,\ldots,B_4$. Since $B_\alpha=M_\alpha$ for $\alpha=1,\ldots,4$, the first four columns of Eq.~\eqref{eq:Truth_table}, with overall signs discarded, determine the finite-state transition matrix
\begin{equation}
    \mathsf{A}
    =
    \begin{pmatrix}
    1 & 1 & 1 & 1 & 0 & 0\\
    1 & 1 & 0 & 0 & 1 & 1\\
    1 & 1 & 1 & 1 & 0 & 0\\
    0 & 0 & 1 & 1 & 1 & 1\\
    1 & 1 & 0 & 0 & 1 & 1\\
    0 & 0 & 1 & 1 & 1 & 1
    \end{pmatrix}.
\end{equation}
Its entry $\mathsf{A}_{\alpha\beta}$ counts the number of letters $B_j$ that map the class of $M_\alpha$ to that of $M_\beta$. Consequently,
\begin{equation}
    \mathbf{x}_{2L}
    = \mathbf{x}_{2L-2}\mathsf{A}.
\end{equation}
The initial vector is
\begin{equation}
    \mathbf{x}_2
    = \begin{pmatrix}1 & 1 & 1 & 1 & 0 & 0\end{pmatrix},
\end{equation}
since each of the four two-bond configurations corresponds to one of $B_1,\ldots,B_4$. The exponentially many configurations are thus counted through an evolution involving only six states.

The absolute traces define the weight vector
\begin{equation}
    \mathbf{v}
    =
    \begin{pmatrix}
    2\\0\\0\\1\\1\\0
    \end{pmatrix},
\end{equation}
and hence
\begin{equation}
    N_{2L}
    = \mathbf{x}_{2L}\mathbf{v}
    = \mathbf{x}_2\mathsf{A}^{L-1}\mathbf{v}.
\end{equation}
In particular,
\begin{equation}
    N_2=3,\qquad N_4=11,\qquad N_6=43.
\end{equation}

The characteristic polynomial of $\mathsf{A}$ is
\begin{equation}
    \chi_{\mathsf{A}}(\lambda)
    = \lambda^3(\lambda-1)^2(\lambda-4).
\end{equation}
Direct multiplication also verifies the stronger identity
\begin{equation}
    \mathsf{A}^3-5\mathsf{A}^2+4\mathsf{A}=0.
\end{equation}
Multiplying this identity on the left by $\mathbf{x}_{2L-2}$ and on the right by $\mathbf{v}$ gives
\begin{equation}
    N_{2L+4}-5N_{2L+2}+4N_{2L}=0,
    \qquad L\geq2.
\end{equation}
The same relation holds for $L=1$, as verified by the initial counts above. Equivalently,
\begin{equation}
    N_{2L+2}
    = 5N_{2L}-4N_{2L-2},
    \qquad L\geq2.
\end{equation}

The Jacobsthal numbers are defined by~\cite{weisstein_jacobsthal}
\begin{equation}
    J_{n+2}=J_{n+1}+2J_n,
    \qquad J_0=0,\quad J_1=1.
\end{equation}
Their odd-indexed subsequence satisfies
\begin{equation}
    J_{2L+3}
    = 5J_{2L+1}-4J_{2L-1}.
\end{equation}
Since $J_3=3$ and $J_5=11$, the initial conditions and recurrence agree with those of the zero-mode count. We therefore obtain
\begin{equation}
    N_{2L}
    = J_{2L+1}
    = \frac{2^{2L+1}+1}{3}.
\end{equation}

The transfer-matrix approach can also be adapted to open boundary conditions and odd-length chains. For open boundaries, the trace is replaced by a contraction with boundary vectors that implement the allowed endpoint states. For odd-length chains, one transfer matrix remains unpaired; its contribution can be incorporated into the initial or terminal weights of the counting procedure. In either case, the boundary weights and finite-state closure must be determined for the corresponding transfer-matrix expression.

\section{Construction of states without defects}\label{app:proof}
We use the transfer-matrix formalism to show that every sector of a periodic chain containing an even number of walls admits exactly two defect-free product states. We also provide an explicit construction of these states.

For a bond with charge $w_j=\pm1$, we define the transfer matrix by
\begin{equation}
T^{(w_j)}_{ki} =
\begin{cases}
1, & \text{if the two-site state $|ki\rangle$ is compatible with $w_j$},\\
0, & \text{otherwise},
\end{cases}
\end{equation}
where $k,i\in\{x,y,z\}$ label the states on sites $j$ and $j+1$, respectively. For periodic boundary conditions, the dimension of the sector specified by $w=\{w_1,w_2,\dots,w_L\}$ is
\begin{equation}
D_w = \operatorname{Tr}\left[
T^{(w_1)}T^{(w_2)}\cdots T^{(w_L)}
\right].
\end{equation}
In the ordered basis $\{x,y,z\}$, the transfer matrices are
\begin{equation}
T^{(+)} =
\begin{pmatrix}
0 & 1 & 1\\
1 & 0 & 0\\
0 & 1 & 1
\end{pmatrix},
\qquad
T^{(-)} =
\begin{pmatrix}
1 & 0 & 0\\
0 & 1 & 1\\
1 & 0 & 0
\end{pmatrix}.
\end{equation}

Defect-free states contain only $|x\rangle$ and $|y\rangle$ states. Restricting the transfer matrices to these two local states gives
\begin{equation}
T^{(+)}_{x,y} =
\begin{pmatrix}
0 & 1\\
1 & 0
\end{pmatrix}
= \sigma_x,
\qquad
T^{(-)}_{x,y} =
\begin{pmatrix}
1 & 0\\
0 & 1
\end{pmatrix}
= \mathbb{I}_2,
\end{equation}
where $\sigma_x$ is the Pauli $x$ matrix. Since each wall corresponds to a bond with $w_j=+1$, the number of defect-free states in a sector with $N_{\mathrm{w}}$ walls is
\begin{equation}
D_w^{(0)}
= \operatorname{Tr}\left[
T^{(w_1)}_{x,y}\cdots T^{(w_L)}_{x,y}
\right]
= \operatorname{Tr}\left(\sigma_x^{N_{\mathrm{w}}}\right).
\end{equation}
Using $\sigma_x^2=\mathbb{1}_2$, we obtain
\begin{equation}
D_w^{(0)} =
\begin{cases}
2, & N_{\mathrm{w}}\ \text{even},\\
0, & N_{\mathrm{w}}\ \text{odd}.
\end{cases}
\end{equation}
Thus, every even-wall sector contains exactly two defect-free product states, irrespective of the arrangement of its bond charges, whereas odd-wall sectors contain none.

The two states can be constructed directly from the bond configuration. Choose the state on the first site to be either $x$ or $y$, and then traverse the chain using the rule
\begin{equation}
s_{j+1} =
\begin{cases}
s_j, & w_j=-1,\\
\bar{s}_j, & w_j=+1,
\end{cases}
\end{equation}
where $s_j\in\{x,y\}$, $\bar{x}=y$, and $\bar{y}=x$. A bond with charge $-1$ preserves the state, while a bond with charge $+1$ exchanges $x$ and $y$. For an even number of walls, the state undergoes an even number of exchanges around the chain, ensuring compatibility with the periodic boundary condition $s_{L+1}=s_1$. Each choice of $s_1$ therefore uniquely determines a defect-free state, and the two states are related by exchanging $x$ and $y$ on every site.

This construction can equivalently be viewed as a two-state finite-state machine, in which the bond charges specify the transition rules between the local states $|x\rangle$ and $|y\rangle$~\cite{FSM_book}.

\section{K-wall Numerics}
\label{app:kwallnumerics}

\begin{figure*}[t]
    \centering
    \includegraphics[width=\textwidth]{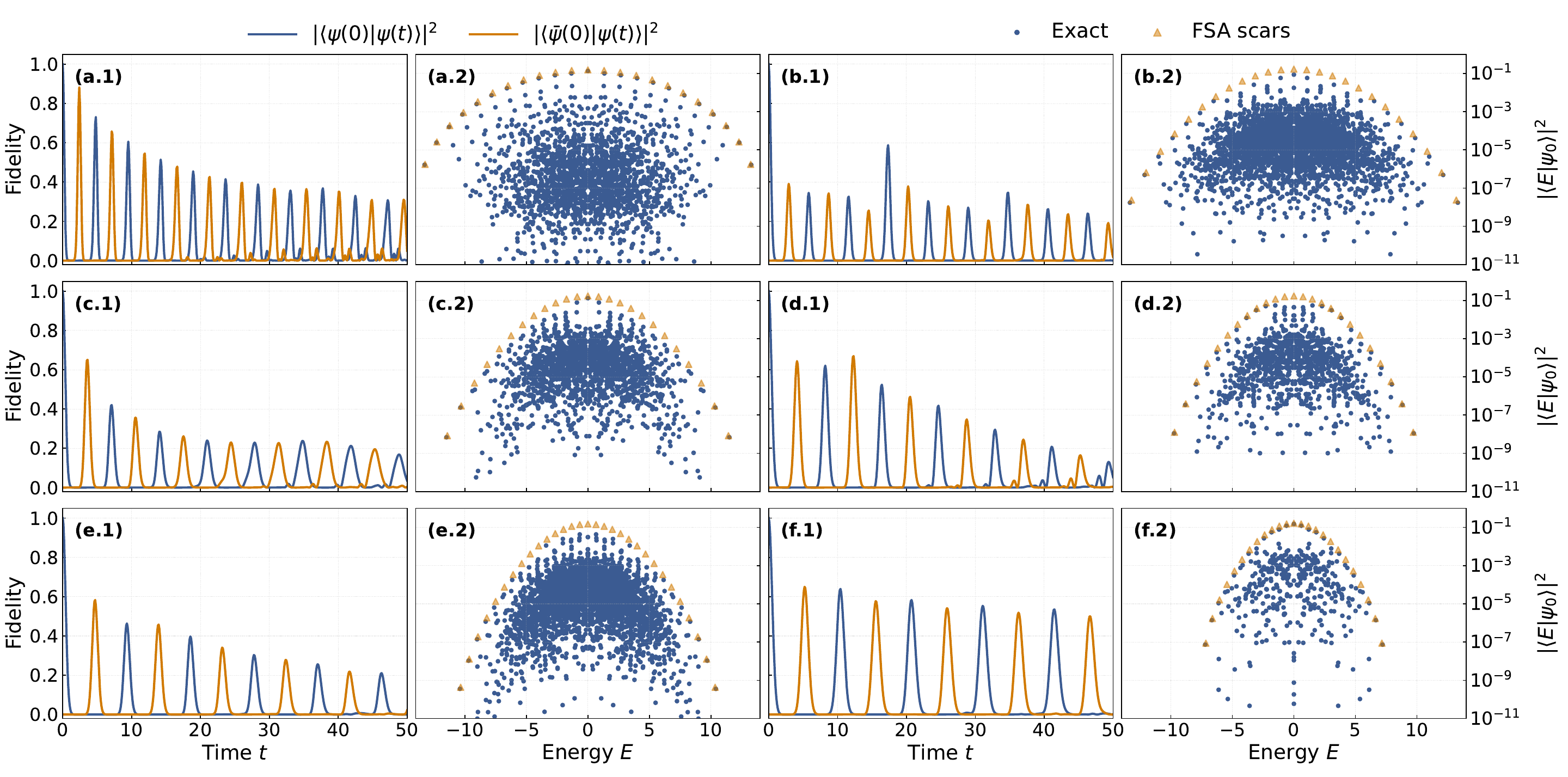}
    \caption{
        Fidelity dynamics and eigenstate overlap spectra for the different wall sectors.
        Panels (a.1) and (a.2) show the fidelity dynamics and overlap spectrum,
        respectively, for the 0-wall sector (PXP) for $L=22$ (dim=39603);
        (b.1) and (b.2) correspond to the 1-wall sector for $L=24$
        (dim=39202); (c.1) and (c.2) to the 2-wall sector for $L=24$
        (dim=14159); (d.1) and (d.2) to the 3-wall sector for $L=24$
        (dim=5778); (e.1) and (e.2) to the 4-wall sector for $L=30$
        (dim=19602); and (f.1) and (f.2) to the 5-wall sector for $L=24$
        (dim=1442).
    }
    \label{app:fig:six_wall_diagnostics}
\end{figure*}
We present numerical evidence for scarred dynamics in several sectors specified by periodic $k$-wall bond-charge configurations. Fig.~\ref{app:fig:six_wall_diagnostics} shows fidelity revivals from the corresponding sector-Neel states (blue curves). We also plot the overlap of the time-evolved state with the sector-anti-Neel state, denoted by $|\bar{\Psi}(0)\rangle$. The alternating peaks in these two quantities indicate coherent transfer between the sector-Neel and sector-anti-Neel states, analogous to the dynamics observed in the PXP model~\cite{Zlatko_Hypergrid_origin}.

The quality of the fidelity revivals varies across sectors. This variation may partly reflect differences in sector dimensions: at fixed system size $L$, the Hilbert-space dimension decreases as $k$ increases. Direct comparisons between sectors therefore require care, and the available results do not establish a general trend in revival quality as a function of $k$.

The revivals persist over the system sizes examined. As a representative example, we consider the $3$-wall periodic sector, whose sector-Neel state is
\begin{equation}
    |N\rangle_w = |xyyyyxxxx\cdots\rangle.
\end{equation}
To assess whether the observed revivals simply reflect the small dimension of this sector, we compare the dynamics from this state with those from an ensemble of randomly sampled product states in the same sector. We also examine their overlaps with the energy eigenstates.

\begin{figure}
    \centering
    \includegraphics[width=0.8\linewidth]{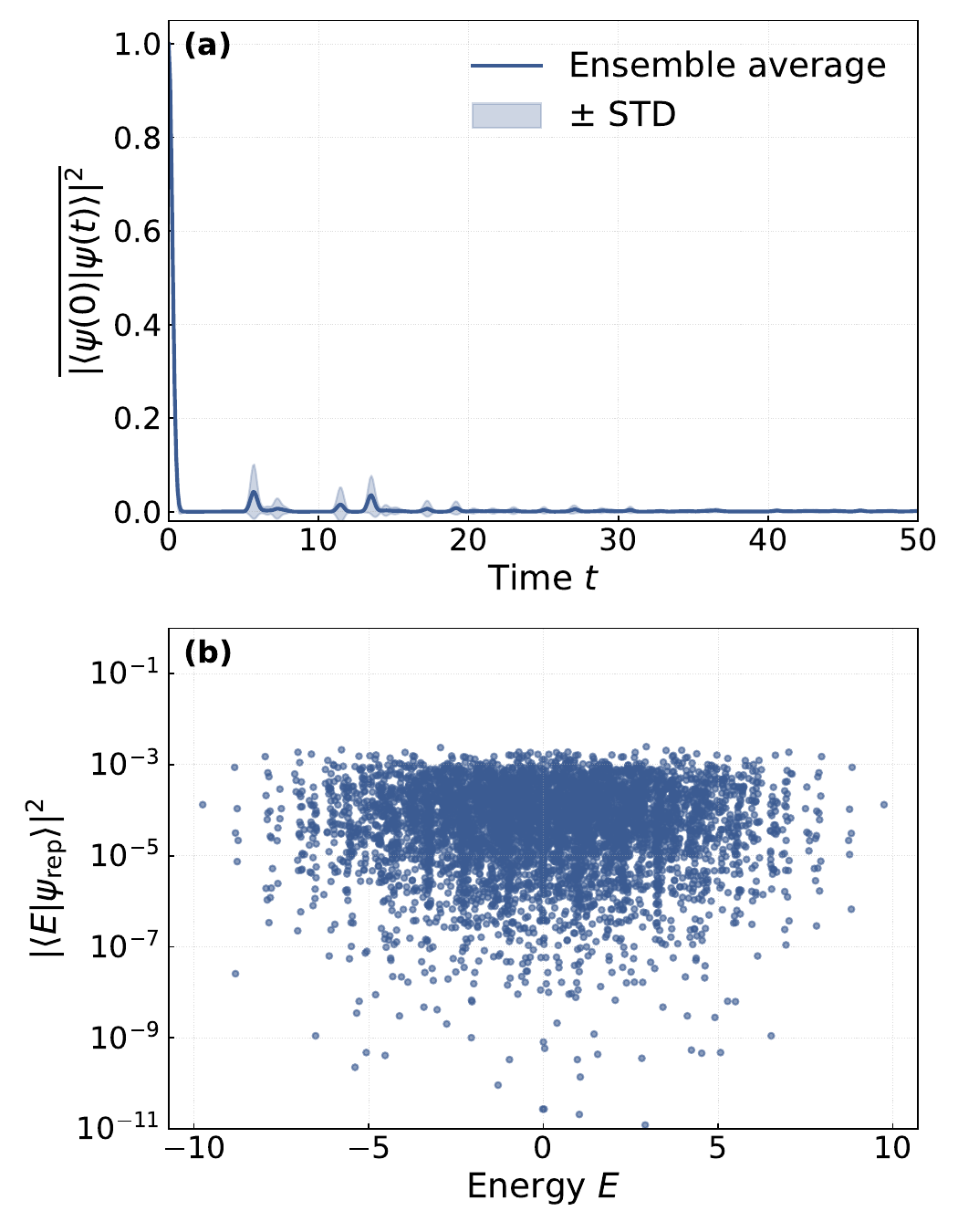}
    \caption{For the $3$-wall sector with $N$=100 random product states, (a) the ensemble-averaged fidelity (where STD stands for Standard Deviation) and (b) the overlap spectrum for a representative state}
    \label{app:fig:random_product_diaganostics}
\end{figure}

For each randomly sampled product state, we compute the fidelity as a function of time and then average over the ensemble. Sampling multiple states reduces the dependence on a particular initial state that might accidentally have appreciable overlap with the scarred eigenstates. The ensemble-averaged fidelity, shown in Fig.~\ref{app:fig:random_product_diaganostics}(a), exhibits no pronounced revivals. Furthermore, the overlap spectrum of a representative state from the ensemble, shown in Fig.~\ref{app:fig:random_product_diaganostics}(b), lacks a discernible tower-like structure. These results support the interpretation that the revivals from the sector-Neel state arise from its particular spectral properties, rather than from the sector dimension alone.

In contrast, the overlap spectrum of the sector-Neel state reveals a sequence of eigenstates with enhanced overlaps relative to nearby eigenstates in the same energy window. Their approximately uniform energy spacing supports coherent rephasing and the resulting fidelity revivals. The revival quality is therefore expected to depend on both the distribution of spectral weight among these eigenstates and deviations from uniform energy spacing. A detailed quantitative analysis of these effects remains an open direction.

\section{k-Bound wall Numerics data}
\label{app:boundwallnumerics}

\begin{figure*}[t]
    \centering
    \includegraphics[width=\textwidth]{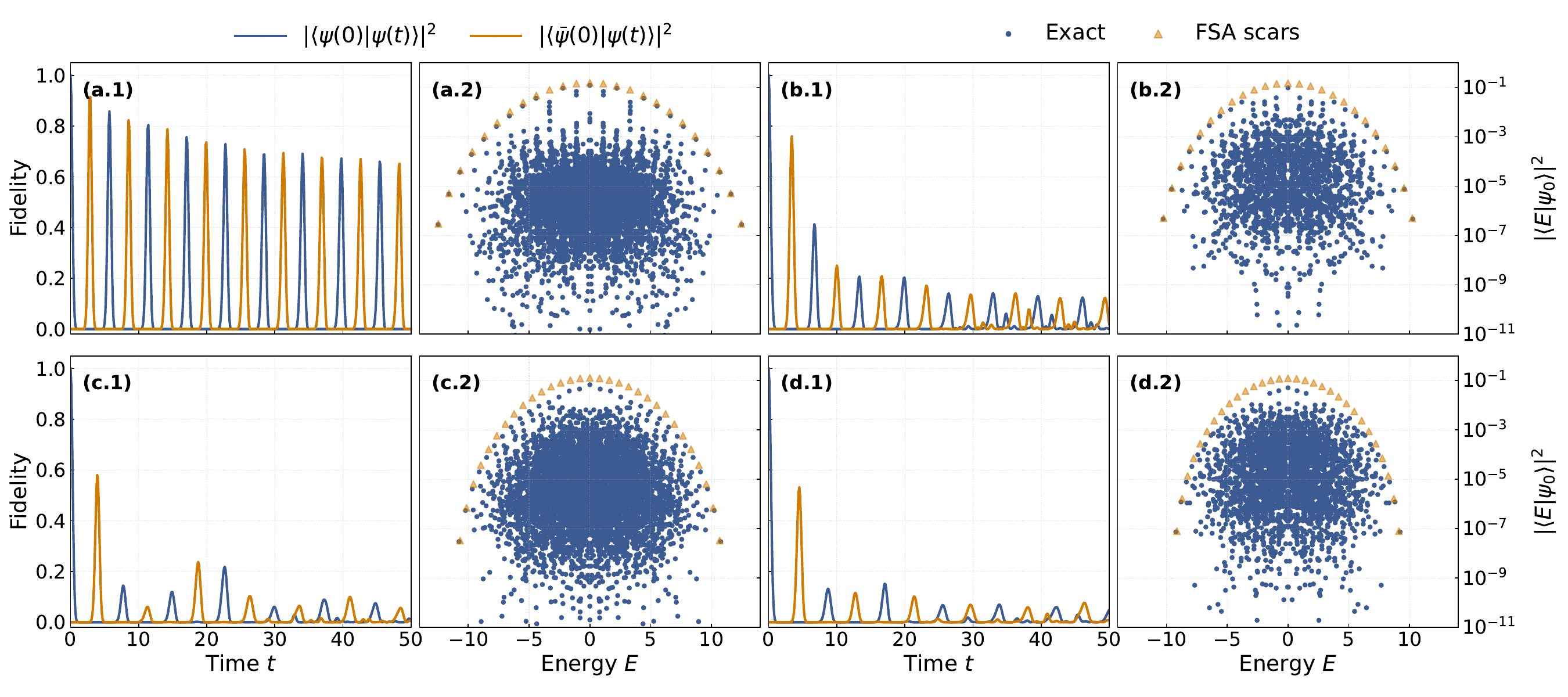}
    \caption{
        Fidelity dynamics and eigenstate overlap spectra for the different wall sectors.
        Panels (a.1) and (a.2) show the fidelity dynamics and overlap spectrum,
        respectively, for the 1-boundwall sector for $L=24$ (dim=37634);
        (b.1) and (b.2) correspond to the 2-boundwall sector for $L=24$
        (dim=12098); (c.1) and (c.2) to the 3-boundwall sector for $L=30$
        (dim=39202); (d.1) and (d.2) to the 4-boundwall sector for $L=30$
        (dim=15127)
    }
    \label{app:fig:4_wall_diagnostics}
\end{figure*}
We examine scarred dynamics in the class of $k$-boundwall sectors. As in the preceding appendix, we find fidelity revivals from the corresponding sector-Neel states. Within the cases examined, however, the revival quality decreases as $k$ increases. This trend is illustrated in Fig.~\ref{app:fig:4_wall_diagnostics}: panels (a) and (c) correspond to the $1$- and $3$-boundwall sectors, respectively, which have comparable Hilbert-space dimensions but markedly different revival amplitudes. A similar reduction is observed when comparing the $2$- and $4$-boundwall sectors.

\begin{figure}
    \centering
    \includegraphics[width=0.8\linewidth]{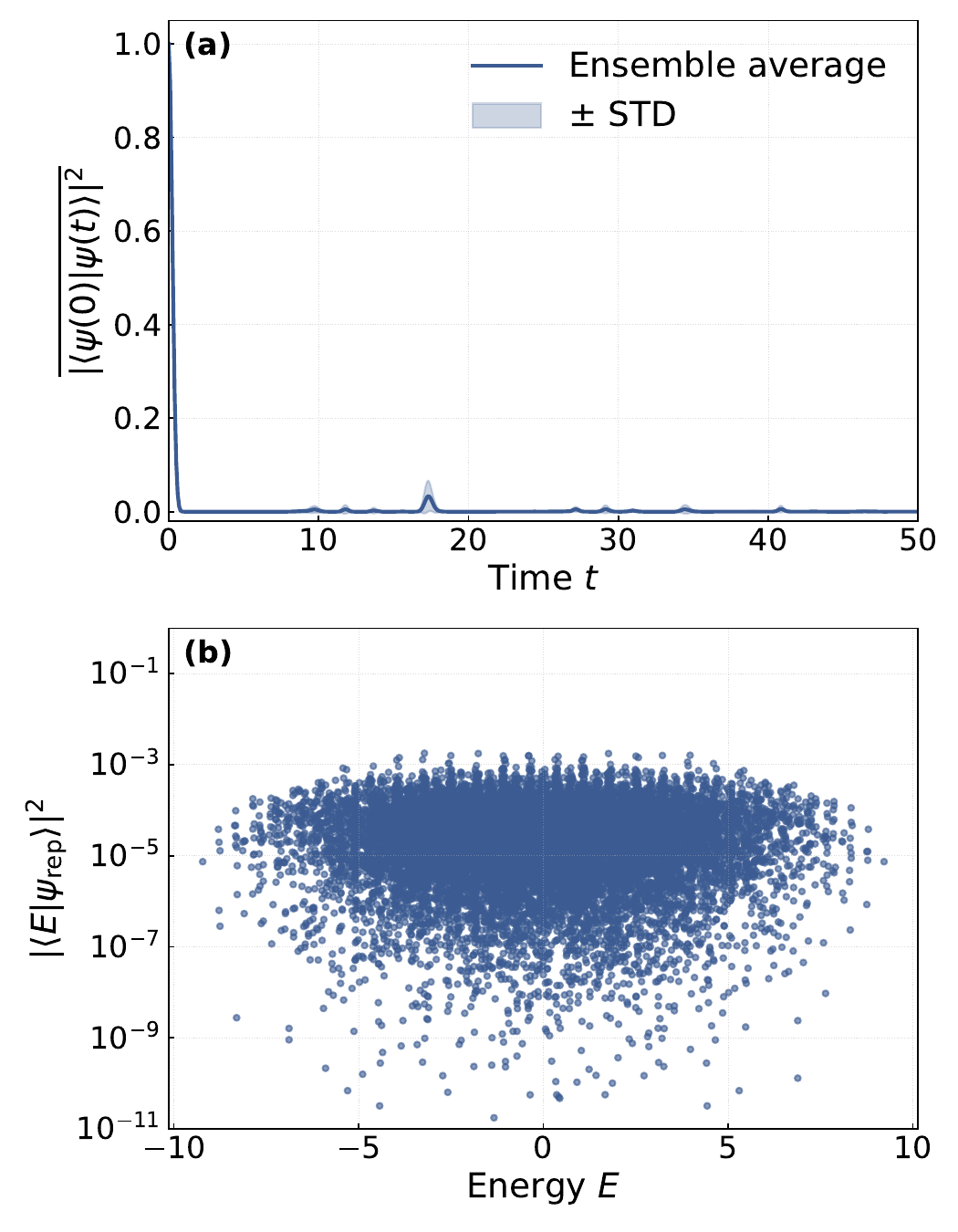}
    \caption{Random-product-state diagnostics for the $4$-boundwall sector at $L=30$: (a) fidelity dynamics averaged over an ensemble of $N=100$ random product states and (b) the corresponding eigenstate overlap spectrum.}
    \label{app:fig:random_4b_diaganostics}
\end{figure}

Although the revivals weaken, they remain discernible. To assess whether they arise solely from the small sector dimension, we examine quench dynamics from an ensemble of random product states in the $4$-boundwall sector. As shown in Fig.\ref{app:fig:random_4b_diaganostics}(a), the ensemble-averaged fidelity exhibits no pronounced revivals, and the corresponding eigenstate overlap spectrum lacks a discernible tower-like structure. These results support the interpretation that the revivals from the sector-Neel state depend on its particular spectral properties, rather than on the Hilbert-space dimension alone.

\begin{figure*}[t]
    \centering
    \includegraphics[width=\textwidth]{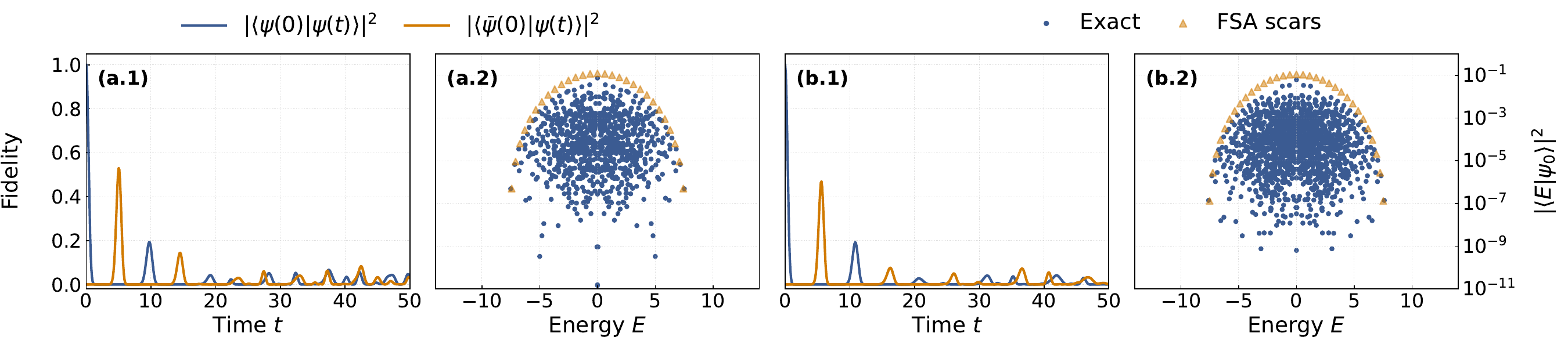}
    \caption{Fidelity dynamics and eigenstate overlap spectra for the $5$- and $6$-boundwall sectors. Panels (a) and (b) show the fidelity dynamics and overlap spectrum, respectively, for the $5$-boundwall sector at $L=28$, with sector dimension $3842$. Panels (c) and (d) show the corresponding results for the $6$-boundwall sector at $L=32$, with sector dimension $6239$.}
    \label{app:fig:56b_wall}
\end{figure*}

The reduction in revival quality is also apparent in the $5$- and $6$-boundwall sectors, shown in Fig.~\ref{app:fig:56b_wall}. Alongside weaker revivals, the overlap spectra show reduced spectral weight on the distinguished eigenstates associated with scarring.

Despite the weaker revivals, the sector-Neel overlap spectra retain sequences of eigenstates with enhanced overlaps relative to nearby eigenstates in the same energy window. This structure motivates exploring whether a suitable local unitary transformation of the initial state could increase its overlap with the scarred subspace and improve the revivals, analogous to the squeezing construction studied in the PXP model~\cite{Hannes_2022_scars_as_scatteringproblem}. Establishing whether such a transformation exists, and quantitatively characterizing the dependence of revival quality on $k$, require further investigation.

\section{Mixed-sector Numerics}\label{app:mixedsectornumerics}
We examine sectors with periodic bond configurations containing isolated walls and bound pairs of walls. An isolated wall consists of a single $w_j=+1$ bond flanked by $w_j=-1$ bonds, whereas a bound pair consists of two consecutive $w_j=+1$ bonds flanked by $w_j=-1$ bonds. We consider system sizes commensurate with the repeated bond motif and require an even total number of walls, ensuring the existence of the sector-Neel states.

\begin{figure*}[t]
    \centering
    \includegraphics[width=\textwidth]{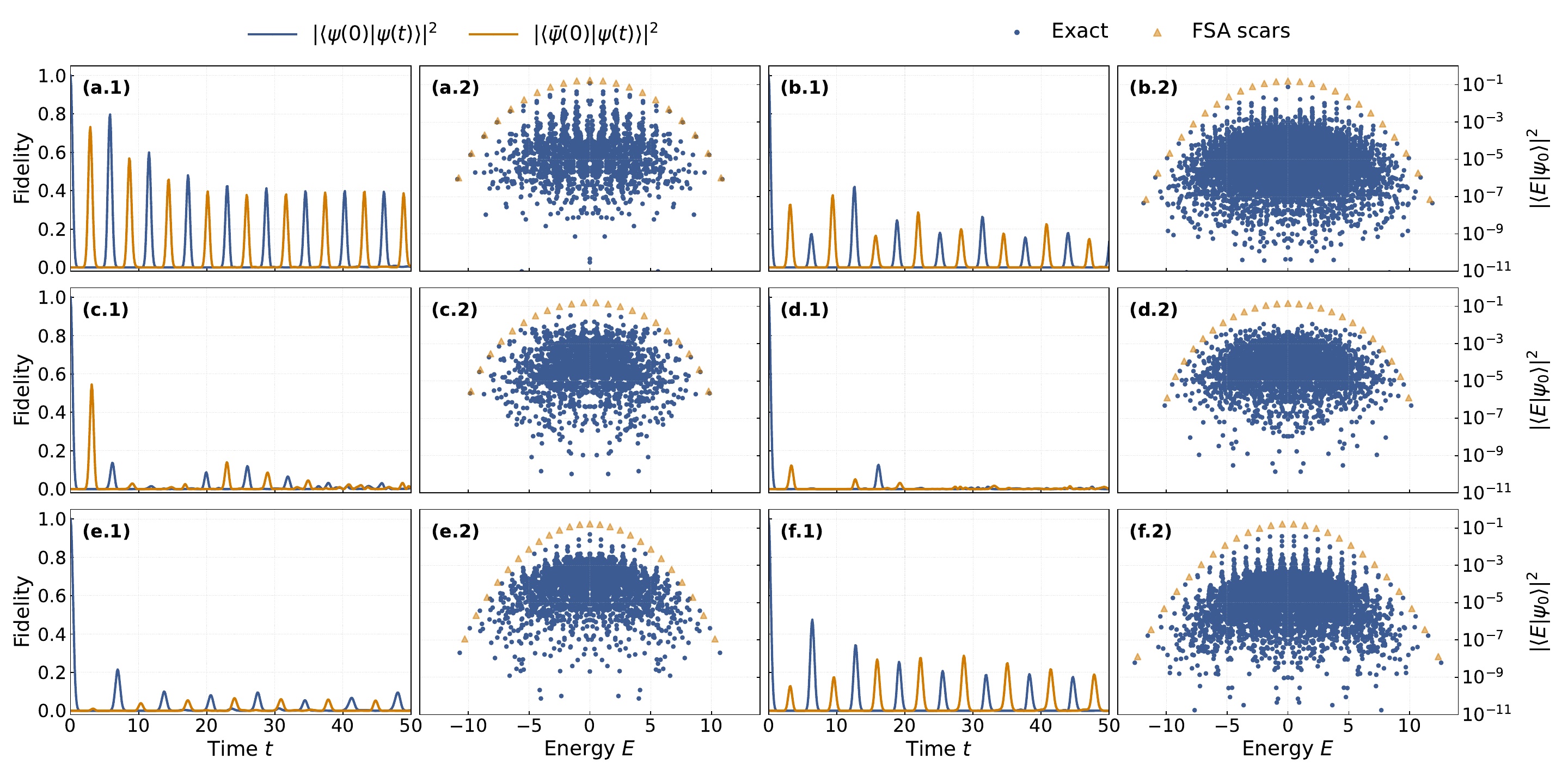}
    \caption{Fidelity dynamics and eigenstate overlap spectra for sectors defined by repeated bond motifs. In each panel pair, the first panel shows the fidelity dynamics and the second shows the overlap spectrum. Panels (a.1) and (a.2) correspond to the motif $\{1,1,-1,1,-1\}$ at $L=20$, with sector dimension $6887$; (b.1) and (b.2) correspond to the motif $\{1,1,-1,-1,1,-1\}$ at $L=24$, with dimension $21314$; (c.1) and (c.2) correspond to the motif $\{1,1,-1,-1,1,1,-1\}$ at $L=21$, with dimension $5778$; (d.1) and (d.2) correspond to the motif $\{1,1,-1,-1,-1,1,1,-1\}$ at $L=24$, with dimension $10582$; (e.1) and (e.2) correspond to the motif $\{1, -1, 1, -1, -1, -1\}$ at $L=24$, with dimension $9048$; (e.1) and (e.2) correspond to the motif $\{1, -1, 1, -1, -1\}$ at $L=25$, with dimension $30248$ }
    \label{app:fig:mixed4_wall_diagnostics}
\end{figure*}

We study four periodic bond motifs:
\begin{align}
    \text{(a)}\quad & \{1,1,-1,1,-1\}, \\
    \text{(b)}\quad & \{1,1,-1,-1,1,-1\}, \\
    \text{(c)}\quad & \{1,1,-1,-1,1,1,-1\}, \\
    \text{(d)}\quad & \{1,1,-1,-1,-1,1,1,-1\}.
\end{align}
Motifs (a) and (b) contain both isolated walls and bound pairs, while motifs (c) and (d) contain bound pairs separated by unequal intervals. As shown in Fig.~\ref{app:fig:mixed4_wall_diagnostics}, the fidelity revivals vary substantially across these configurations. They become progressively less pronounced in the examples shown, with motif (d) exhibiting no appreciable revivals. Since both the system sizes and sector dimensions differ, these results illustrate the sensitivity of scarred dynamics to the bond configuration rather than establish a general trend with motif complexity.

These observations raise the question of which properties of a sector determine whether its sector-Neel state exhibits scarred dynamics. A systematic answer lies beyond the scope of this work. One possible approach is to examine whether the dynamics generated from this initial state admit an approximate Lie-algebraic description, and whether deviations from algebraic closure account for the variation in revival quality, following the broken Lie-algebra framework of Ref.~\cite{Weakly_broken_lie_algebra_pxp}.

\begin{figure}
    \centering
    \includegraphics[width=0.8\linewidth]{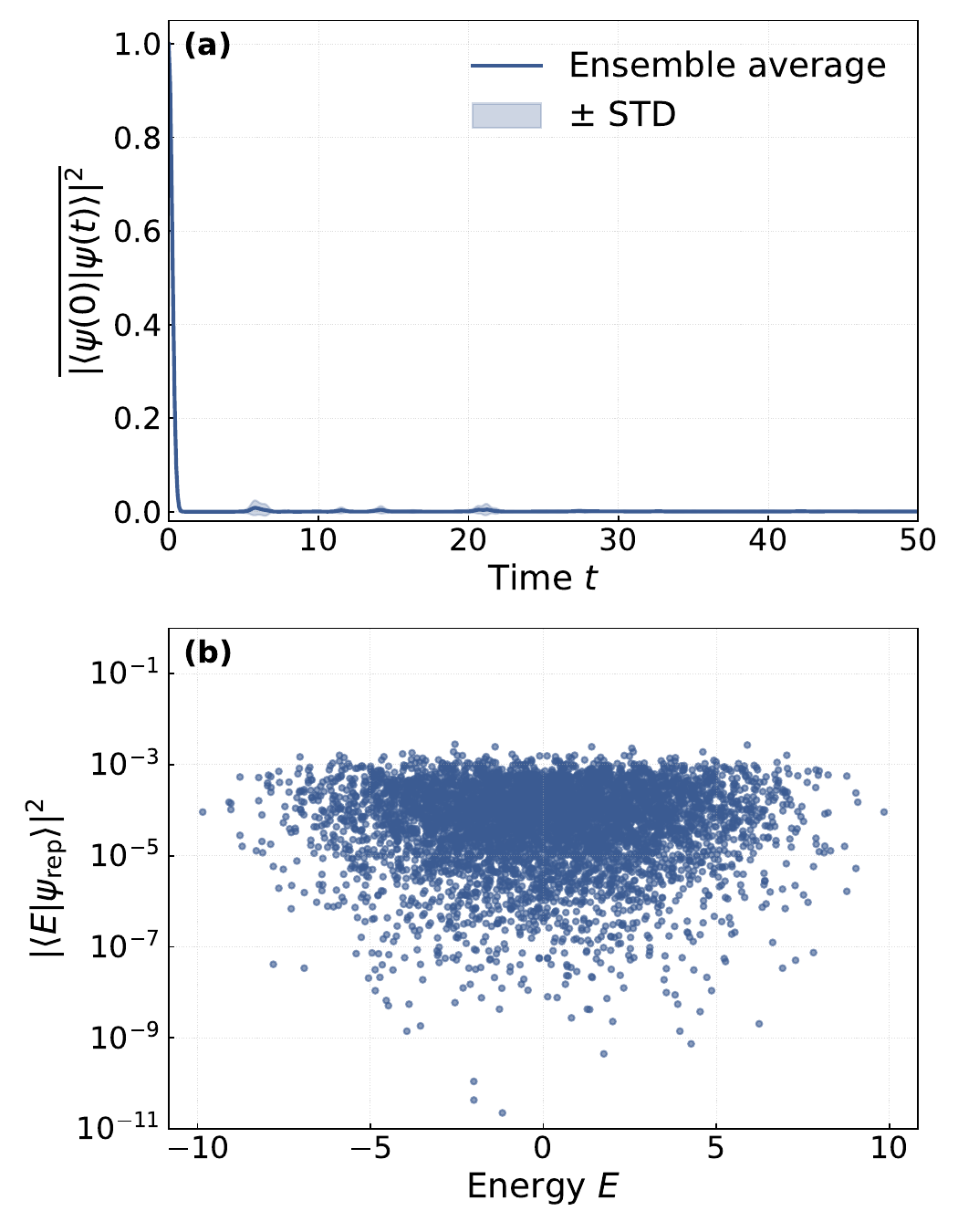}
    \caption{Random-product-state diagnostics for the sector defined by the repeated motif $\{1,1,-1,-1,1,1,-1\}$ at $L=21$. (a) Fidelity averaged over an ensemble of $N=100$ random product states. (b) Eigenstate overlap spectrum of a representative state from the ensemble, showing no discernible tower-like structure.}
    \label{app:fig:random_diag_mixed1}
\end{figure}

To further assess the dependence of the revivals on the initial state, we examine an ensemble of random product states in the sector defined by motif (c). As shown in Fig.~\ref{app:fig:random_diag_mixed1}, the ensemble-averaged fidelity exhibits no pronounced revivals, while the overlap spectrum of a representative state lacks a discernible tower-like structure. These results support the interpretation that the revivals from the sector-Neel state reflect its enhanced overlap with scarred eigenstates, rather than the finite sector dimension alone.

\section{Non-scarred sectors}\label{app:Nonscarred_Sectors}
\begin{figure*}[t]
    \centering
    \includegraphics[width=\textwidth]{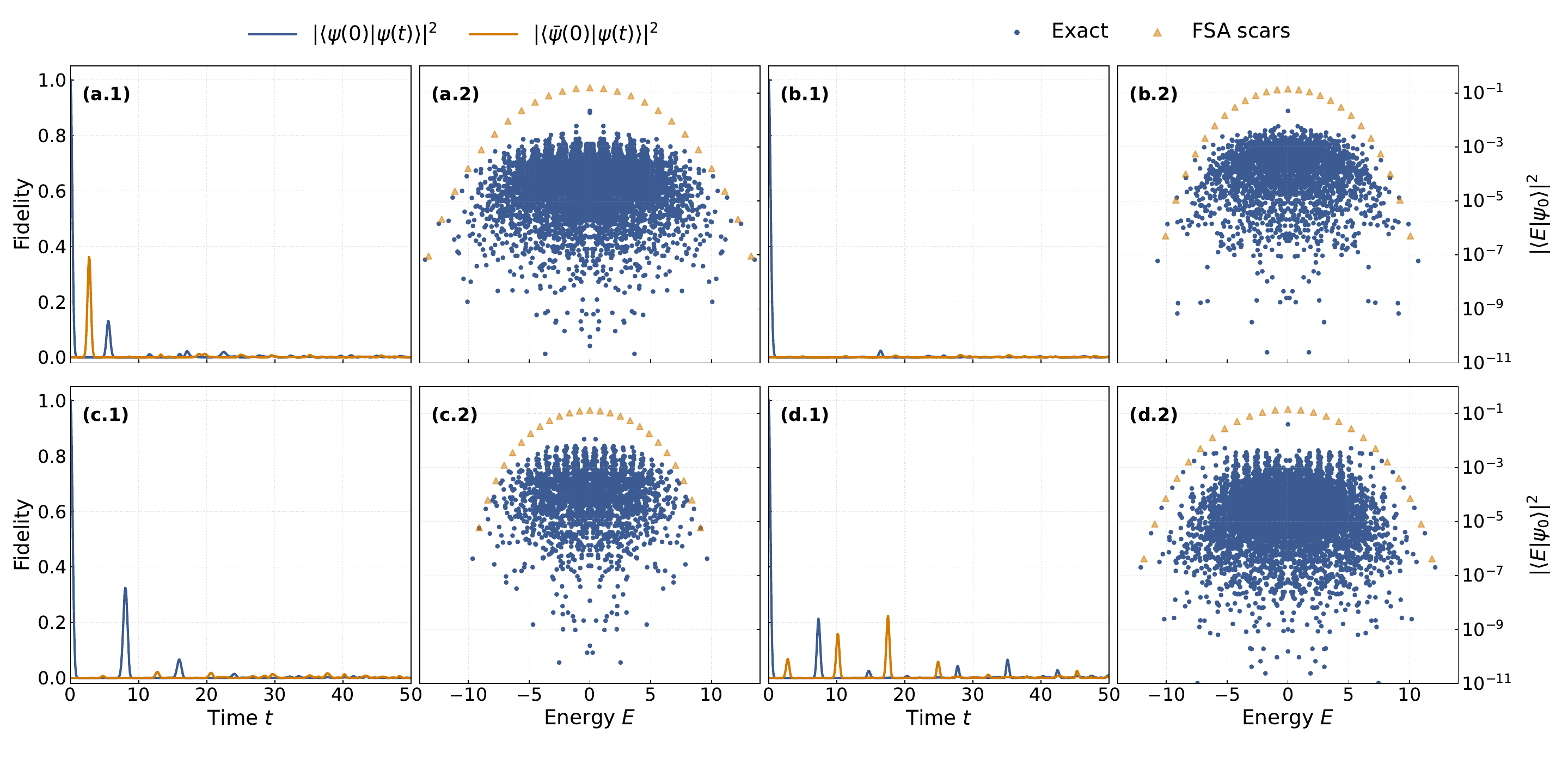}
    \caption{
        Fidelity dynamics and eigenstate overlap spectra for the different wall sectors.
        Panels (a.1) and (a.2) show the fidelity dynamics and overlap spectrum,
        respectively, for the sector $w=\{1, 1, 1, -1, 1, -1\}$ for $L=24$ (dim=49727);
        (b.1) and (b.2) correspond to the sector $w=\{1, 1, 1, -1, -1, -1\}$ for $L=24$
        (dim=10402); (c.1) and (c.2) correspond to the sector $w=\{1, 1, 1, 1, -1, -1, -1, -1\}$ for $L=24$
        (dim=6802); (d.1) and (d.2) to the sector $w=\{1, 1, 1, 1, -1, -1\}$ for $L=24$
        (dim=27887)
    }
    \label{app:fig:nonscar_diagnostics}
\end{figure*}

We present sectors that exhibit no sustained fidelity revivals from their sector-Neel states, as shown in Fig.~\ref{app:fig:nonscar_diagnostics}. Some cases nevertheless display an isolated revival peak. For example, the sector defined by the repeated bond motif $\{1,1,1,-1,-1,-1\}$ exhibits a single peak in Fig.~\ref{app:fig:nonscar_diagnostics}(c), whereas its counterpart shows no appreciable revivals. This contrast suggests that the isolated peak may reflect a finite-size effect or dynamics arising from a mechanism distinct from the slider picture studied here.

\end{document}